\documentclass[twocolumn,aps,pra,longbibliography,superscriptaddress]{revtex4-2}
\usepackage{amsmath,latexsym}
\usepackage{xcolor}
\usepackage[%
  colorlinks=true,
  urlcolor=blue,
  linkcolor=blue,
  citecolor=blue
]{hyperref}
\usepackage{etoolbox}
\usepackage{breqn}
\usepackage{mathrsfs,verbatim,mathtools,graphicx,amsmath, amssymb, bm, epstopdf,dsfont, enumerate}
\makeatletter
\let\cat@comma@active\@empty
\makeatother

\begin{document}

\title{A classical model of spontaneous parametric down-conversion}

\author{Girish Kulkarni}
\email{girishkulkarni@protonmail.com}
\affiliation{Department of Physics, University of Ottawa, Ottawa, Ontario K1N 6N5, Canada}
\author{Jeremy Rioux}
\affiliation{Department of Physics, University of Ottawa, Ottawa, Ontario K1N 6N5, Canada}
\author{Boris Braverman}
\email{Present address: QuEra Computing, Boston, MA 02135, USA}
\affiliation{Department of Physics, University of Ottawa, Ottawa, Ontario K1N 6N5, Canada}
\author{Maria V. Chekhova}
\affiliation{Max Planck Institute for the Science of Light, Staudtstr. 2, 91058 Erlangen, Germany}
\affiliation{University of Erlangen-Nuremberg, Staudtstr. 7/B2, 91058 Erlangen, Germany}
\author{Robert W. Boyd}
\affiliation{Department of Physics, University of Ottawa, Ottawa, Ontario K1N 6N5, Canada}
\affiliation{The Institute of Optics, University of Rochester, Rochester, New York 14627, USA}
\date{\today}
\begin{abstract}
We model spontaneous parametric down-conversion (SPDC) as classical difference frequency generation (DFG) of the pump field and a hypothetical stochastic ``vacuum'' seed field. We analytically show that the second-order spatiotemporal correlations of the field generated from the DFG process replicate those of the signal field from SPDC. Specifically, for low gain, the model is consistent with the quantum calculation of the signal photon's reduced density matrix; and for high gain, the model's predictions are in good agreement with our experimental measurements of the far-field intensity profile, orbital angular momentum spectrum, and wavelength spectrum of the SPDC field for increasing pump strengths. We further theoretically show that the model successfully captures second-order SU(1,1) interference and induced coherence effects in both gain regimes. Intriguingly, the model also correctly predicts the linear scaling of the interference visibility with object transmittance in the low-gain regime -- a feature that is often regarded as a quintessential signature of the nonclassicality of induced coherence. Our model may not only lead to novel fundamental insights into the classical-quantum divide in the context of SPDC and induced coherence, but can also be a useful theoretical tool for numerous experiments and applications based on SPDC.
\end{abstract}
\maketitle
\section{Introduction} 
Spontaneous parametric down-conversion (SPDC) is a nonlinear optical phenomenon in which an incident field known as pump interacts with a non-centrosymmetric medium to produce a pair of fields known as signal and idler \cite{klyshko1967jetp,burnham1970prl,boyd2020}. In the quantum paradigm, the interaction can be modeled as a series of infinitely-many contributing processes, where the $n$'th order contribution physically corresponds to $n$ pump photons getting annihilated to produce $n$ signal photons and $n$ idler photons simultaneously \cite{couteau2018tf}. The strength of the interaction, referred to as gain, is directly proportional to the pump amplitude \cite{klyshko1988}. For typical continuous-wave and low-power pulsed pump lasers, the interaction is weak and can be approximated by the dominant first-order contribution alone, which yields an entangled two-photon signal-idler state \cite{hong1985pra, keller1997pra, grice1997pra}. In this low-gain regime, which is characterized by a linear growth of the generated fields with respect to the pump amplitude, SPDC sources are ubiquitously employed in fundamental quantum optics experiments \cite{hong1987prl,shih1988prl,zou1991prl,weihs1998prl} and optical quantum technologies \cite{bouwmeester1997nature, jennewein2000prl,moreau2019natrevphys}. On the other hand, for high-power pulsed pump lasers, the interaction can be strong and may comprise of significant higher-order contributions, thereby producing a bright multiphoton entangled state of the signal and idler fields \cite{brambilla2004pra, agafonov2010pra}. In this high-gain regime, which is characterized by an exponential growth of the generated fields with respect to the pump amplitude, SPDC has significant potential for applications such as sub-shot noise imaging \cite{jedrkiewicz2004prl,brida2010natphot} and generation of nonclassical states of light \cite{harder2016prl,iskhakov2012prl}. 

While the precise form of two-photon state produced in the perturbative low-gain regime has long been analytically derived for a general pump profile \cite{hong1985pra, keller1997pra, grice1997pra}, the theoretical characterization of the non-perturbative high-gain regime has proved to be much more difficult. In order to correctly evaluate higher-order contributions, one must account for the fact that the interaction Hamiltonian may not commute with itself at different times, and consequently, the Schrodinger time-evolution operator must strictly be expanded in terms of the time-ordered Dyson series instead of the usual Taylor series \cite{branczyk2011aip}. However, it was shown that any finite-order truncation of the Dyson series destroys the Gaussian character of the output state \cite{quesada2014pra}, whereas the exact solution can be proved to be a Gaussian state solely from the quadratic form of the interaction Hamiltonian \cite{braunstein2005pra}. Such issues can be avoided by turning to the Heisenberg picture, where one instead obtains a set of coupled differential equations for the evolution of the signal and idler mode operators \cite{klyshko1988, dayan2007pra, wasilewski2006pra, quesada2020pra}. However, these equations have so far been exactly solved only for the special case of a monochromatic plane-wave pump \cite{klyshko1988}, and approximately solved for the cases of a broadband pump with a single spatial mode \cite{dayan2007pra}, and a pump with narrow spectral and angular bandwidth \cite{brambilla2004pra}. For more general cases, numerical solutions that are often computationally intensive have to be sought. Some studies have explored analytical ansatz solutions that assume gain-independent Schmidt modes for the output state \cite{christ2011njp, sharapova2015pra}. However, these solutions have limited applicability as they are unable to account for experimental observations such as broadening of the angular spectrum and wavelength spectrum of the SPDC field with increasing gain \cite{spasibko2012oe, sharapova2020prr}. Thus, a consistent analytical quantum description for high-gain SPDC has still not been achieved for a general pump field.

In view of the aforementioned challenges encountered in performing a general quantum calculation, one can instead ask: is there a {\em classical} analytic description that can consistently explain at least a restricted class of correlations in high-gain SPDC?  For instance, in the low-gain regime, it is known that second-order (in fields) autocorrelations of the individual fields can be modeled classically \cite{pires2011pra, louisell1961pr, schroder1983oqe, gatti1997pra, sivan2020lpr}. A simple and elegant approach in this context was developed in Ref.~\cite{pires2011pra}, where SPDC is modeled as the paraxial propagation of a classical partially-coherent source created in the nonlinear medium by the coupling of the pump field to a classical noise field that simulates the vacuum fluctuations. The second-order correlations derived using this approach were demonstrated to be consistent with the quantum calculation of the reduced one-photon density matrix in the low-gain regime. However, this approach does not appear to admit a straightforward generalization to the high-gain regime as the generated field is a priori assumed to scale linearly with the pump amplitude. Another approach, which has also been neatly summarized in  Ref.~\cite{pires2011pra}, is to model SPDC as a three-wave mixing process between the pump and the vacuum fluctuations \cite{louisell1961pr,schroder1983oqe, gatti1997pra}. In particular, the spatial and spectral profiles of the SPDC field can be derived by solving differential equations involving a stochastic term that represents the vacuum fluctuations \cite{schroder1983oqe, gatti1997pra}, although the origin of coherence is less clear and has been attributed to various specific experimental conditions \cite{trapani1998prl, picozzi2001pre}. However, to our knowledge, this latter approach has not been taken beyond the monochromatic plane-wave pump approximation in the high-gain regime to derive analytic solutions for realistic pump fields with finite spectral and angular bandwidth. Moreover, this latter approach has also not yet been employed for analyzing two-crystal settings such as SU(1,1) interferometers \cite{yurke1986pra,klyshko1993jetplett} and induced coherence experiments \cite{zou1991prl,lemos2014nature}. 

In this work, we adopt the aforementioned latter approach and model SPDC as classical difference frequency generation (DFG) of the pump field and a classical stochastic seed field that simulates the zero-point vacuum fluctuations. In contrast with previous studies, we derive analytical expressions for the second-order spatiotemporal correlations of the generated signal field for a narrow-but-finite-bandwidth pump in low and high gain regimes. The paper is organized as follows: In Section~\ref{theory}, we set up the model and derive the second-order spatiotemporal correlation function of the generated field in low and high gain regimes. In Section~\ref{single-crystal}, we perform experimental measurements of the far-field intensity profile, orbital angular momentum spectrum, and wavelength spectrum for increasing pump strengths and demonstrate their agreement with the model's predictions. In Section~\ref{two-crystal}, we extend the model to analyze SU(1,1) interference and induced coherence effects in both low and high gain regimes. In Section~\ref{conclusions}, we conclude with a summary and outlook. 

\section{Theoretical Model}\label{theory}
\begin{figure}[t!]
\centering
\includegraphics[width=75mm,keepaspectratio]{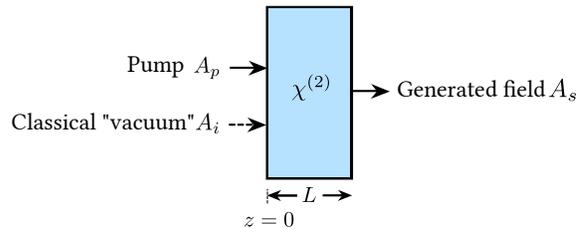}\vspace{-2mm}
\caption{Modeling SPDC as DFG of the pump with a hypothetical classical ``vacuum'' seed field.} \vspace{-2mm}
\label{spdcdfg}
\end{figure}
As depicted in Fig.~\ref{spdcdfg}, we model SPDC as DFG of the pump field $A_{p}$ with a hypothetical classical seed field $A_{i}$ that mimics the effect of zero-point vacuum fluctuations. Our goal is to compute the second-order spatiotemporal correlations of the generated field $A_{s}$. We approximate the fields to be paraxially propagating along the longitudinal $z$-axis. A crystal of length $L$ and second-order nonlinear susceptibility $\chi^{(2)}$ is placed perpendicular to the $z$-axis with its input face defining $z=0$. The Fourier amplitudes (not including propagation phases) for the pump field, classical ``vacuum'', and the generated fields are denoted by stochastic functions $A_{j}({\bm q_{j}},\omega_{j},z)$, where ${\bm q_{j}}$ and $\omega_{j}$ represent the transverse wavevector and frequency coordinates, for $j=p,i,$ and $s$, respectively. In terms of the position vector ${\bm r}\equiv({\bm \rho},z)$, where ${\bm \rho}$ is transverse position, and wavevectors ${\bm k_{j}}\equiv({\bm q_{j}},k_{jz})$, we define the electric field mode functions
\begin{align}\label{fourier}
 \bm{E}_{j}({\bm r},\omega_{j})&=\int\mathrm{d}{\bm q_{j}} A_{j}({\bm q_{j}},\omega_{j},z)\,e^{i({\bm q_{j}}\cdot {\bm \rho}+k_{jz}z)}.
\end{align}
We now assume that the energy depletion of the pump due to the nonlinear interaction is negligible, and consequently, the pump amplitude $A_{p}({\bm q_{p}},\omega_{p})$ is independent of $z$. We emphasize that the pump field can be pulsed or continuous-wave with any arbitrary spectral and spatial profile. The evolution of the other two fields is governed by the wave equation of nonlinear optics (see Eq.~(2.1.23) of Ref.~\cite{boyd2020})
\begin{align}\label{nlowaveqn}
 \nabla^2 \bm{E}_{j}({\bm r},\omega_{j})+\frac{n^2_{j}\omega^2_{j}}{c^2} \bm{E}_{j}({\bm r},\omega_{j})=-\frac{\omega^2_{j}}{\epsilon_{0}c^2} \bm{P}_{NL}({\bm r},\omega_{j}),
\end{align}
where $j=s,i$ and $n_{j}\equiv n_{j}(\omega_{j})$ is the refractive index inside the crystal. Assuming energy conservation, i.e, $\omega_{p}=\omega_{s}+\omega_{i}$, the nonlinear polarization $\bm{P}_{NL}({\bm r},\omega_{j})$ can be written as (see Eq.~(1.5.30) of Ref.~\cite{boyd2020})
\begin{align}\label{nlpol}
 \bm{P}_{NL}({\bm r},\omega_{j})=4\epsilon_{0}d_{\rm eff}\int \mathrm{d}\omega_{p}\,\bm{E}_{p}({\bm r},\omega_{p})\bm{E}_{l}^*({\bm r},\omega_{l}),
\end{align}
where $j=s(i)$ and $l=i(s)$. In our scalar treatment, the quantity $d_{\rm eff}=\chi^{(2)}/2$ is the nonlinear coupling coefficient of the crystal medium, and its dependence on the frequencies of the coupling waves has been ignored. Substituting equations (\ref{fourier}) and (\ref{nlpol}) in (\ref{nlowaveqn}) and making the slowly-varying envelope approximation $\frac{\partial^2}{\partial z^2}A_{j}({\bm q_{j}},\omega_{j},z)\ll k_{jz}\frac{\partial}{\partial z}A_{j}({\bm q_{j}},\omega_{j},z)$ for $j=s,i$, we obtain (see Appendix A for the detailed calculation)
\begin{subequations}\label{dfgeqs}
\begin{align}\notag
 \frac{\partial A_{s}({\bm q_{s}},\omega_{s},z)}{\partial z}&=\frac{2id_{\rm eff}\omega^2_{s}}{k_{sz}c^2}\iint\mathrm{d}\omega_{i}\,\mathrm{d}{\bm q_{i}}\,A_{p}({\bm q_{p}},\omega_{p})\\\label{dfg1}&\hspace{2cm}\times A^*_{i}({\bm q_{i}},\omega_{i},z)e^{i\Delta k_{z}z},\\\notag
\frac{\partial A^*_{i}({\bm q_{i}},\omega_{i},z)}{\partial z}&=\frac{-2id_{\rm eff}\omega^2_{i}}{k_{iz}c^2}\iint\mathrm{d}\omega_{s}\,\mathrm{d}{\bm q_{s}}\,A^*_{p}({\bm q_{p}},\omega_{p})\\\label{dfg2}&\hspace{15mm}\times A_{s}({\bm q_{s}},\omega_{s},z)e^{-i\Delta k_{z}z},
\end{align}
\end{subequations}
where $\Delta k_{z}=k_{pz}-k_{sz}-k_{iz}$ is the longitudinal phase mismatch, and its dependence on ${\bm q_{s}},{\bm q_{i}},\omega_{s},$ and $\omega_{i}$ has been suppressed for brevity. Also, we have assumed that the transverse dimensions of the crystal are much larger than those of the pump beam, which leads to conservation of transverse momentum, i.e, ${\bm q_{p}}={\bm q_{s}}+{\bm q_{i}}$. The above coupled differential equations (\ref{dfgeqs}) describe the nonlinear interaction for arbitrary strengths of the pump field. 
\subsection{Low-gain regime}
When the pump power is low, the nonlinear interaction is weak. In the context of DFG with monochromatic plane waves, it is known that under weak interaction, the longitudinal growth of the seed wave is negligible (see Sec.~2.8 of Ref.~\cite{boyd2020}). Hence, we approximate $A^*_{i}({\bm q_{i}},\omega_{i},z)\approx A^*_{i}({\bm q_{i}},\omega_{i},0)$ and simply integrate Eq.~(\ref{dfg1}) from $z=0$ to $z=L$ to obtain
\begin{align}\notag
 &A_{s}({\bm q_{s}},\omega_{s},L)=\frac{2id_{\rm eff}\omega^2_{s}}{k_{sz}c^2}\iint\mathrm{d}\omega_{i}\,\mathrm{d}{\bm q_{i}}\,A_{p}({\bm q_{p}},\omega_{p})\\\notag&\hspace{3cm}\times A^*_{i}({\bm q_{i}},\omega_{i},0)\int_{0}^{L}\mathrm{d}z\,e^{i\Delta k_{z}z}\\\notag
 &=\frac{d_{\rm eff}L\omega^2_{s}}{k_{sz}c^2}\iint\mathrm{d}\omega_{i}\,\mathrm{d}{\bm q_{i}}\,A_{p}({\bm q_{p}},\omega_{p})A^*_{i}({\bm q_{i}},\omega_{i},0)\\\label{lgspdcfield}&\hspace{15mm}\times\mathrm{sinc}\left(\Delta k_{z}L/2\right)\,\mathrm{exp}\left(i\Delta k_{z}L/2\right),
\end{align}
where $\mathrm{sinc}(x)\equiv (\sin x)/x$. Using the above equation, we can evaluate the second-order correlation function \cite{mandel1995cup}
\begin{align}\notag
&\langle A_{s}({\bm q_{s}},\omega_{s},L)A^*_{s}({\bm q'_{s}},\omega'_{s},L) \rangle=\frac{d_{\rm eff}^2 L^2\omega^2_{s}{\omega'_{s}}^2}{k_{sz}k'_{sz}c^4}\iiiint\mathrm{d}\omega_{i}\,\mathrm{d}{\bm q_{i}}\\\notag&\hspace{3mm}\times\mathrm{d}\omega'_{i}\,\mathrm{d}{\bm q'_{i}}\,\langle A_{p}({\bm q_{s}}+{\bm q_{i}},\omega_{s}+\omega_{i}) A^*_{p}({\bm q'_{s}}+{\bm q'_{i}},\omega'_{s}+\omega'_{i})\rangle\\\notag&\hspace{13mm}\times\langle A_{i}({\bm q'_{i}},\omega'_{i},0) A^*_{i}({\bm q_{i}},\omega_{i},0)\rangle \,\mathrm{sinc}\left(\Delta k_{z}L/2\right)\\\label{lgspdc1}&\hspace{23mm}\times\mathrm{sinc}\left(\Delta k'_{z}L/2\right)\,e^{i\left(\Delta k_{z}-\Delta k'_{z}\right)L/2},
\end{align}
where $\langle \cdots \rangle$ denotes an ensemble average and $\Delta k'_{z}$ is the phase mismatch evaluated for the primed variables. We have assumed that the pump and classical ``vacuum'' fields have no mutual correlation, and as a result, their individual correlation functions factor out separately. 

In order to evaluate the above equation (\ref{lgspdc1}), we must now substitute for $\langle A_{i}({\bm q'_{i}},\omega'_{i},0) A^*_{i}({\bm q_{i}},\omega_{i},0)\rangle$. While it may seem reasonable to assume that different modes of the classical ``vacuum'' are completely uncorrelated, i.e, $\langle A_{i}({\bm q'_{i}},\omega'_{i},0) A^*_{i}({\bm q_{i}},\omega_{i},0)\rangle=C\,\delta({\bm q_{i}}-{\bm q'_{i}})\delta(\omega_{i}-\omega'_{i})$, it is possible to explicitly derive this relation along with the value of the scaling factor $C$ if we make a short detour to quantum theory. We note from the quantized theory of the electromagnetic field that (see Sec.~2.3.3 of Ref.~\cite{ou2017book})
\begin{align}\label{qfieldoperator}
\hat{\bm{E}_{i}}({\bm r},\omega_{i})=i\int \mathrm{d}{\bm q_{i}}\,\sqrt{\frac{\hbar\omega_{i}}{4\pi\epsilon_{0}}}\,\hat{a}({\bm q_{i}},\omega_{i})\,e^{i({\bm q_{i}}\cdot {\bm \rho}+k_{iz}z)},
\end{align}
where $\hat{\bm{E}_{i}}({\bm r},\omega_{i})$ is the quantized operator corresponding to the classical amplitude $\bm{E}_{i}({\bm r},\omega_{i})$ defined in Eq.~(\ref{fourier}) and $\hat{a}({\bm q_{i}},\omega_{i})$ is the annihilation operator for the mode specified by its arguments. Upon comparing equations (\ref{fourier}) and (\ref{qfieldoperator}), it is evident that $A_{i}({\bm q_{i}},\omega_{i},0)$ corresponds to $i\sqrt{\frac{\hbar\omega_{i}}{4\pi\epsilon_{0}}}\,\hat{a}({\bm q_{i}},\omega_{i})$. Now for $A_{i}({\bm q_{i}},\omega_{i},0)$ to effectively simulate the zero-point vacuum fluctuations, we must have $\langle A_{i}({\bm q'_{i}},\omega'_{i},0) A^*_{i}({\bm q_{i}},\omega_{i},0)\rangle=\frac{\hbar\sqrt{\omega_{i}\omega'_{i}}}{4\pi\epsilon_{0}}\langle \mathrm{vac}|\hat{a}_{i}({\bm q'_{i}},\omega'_{i}) \hat{a}^{\dagger}_{i}({\bm q_{i}},\omega_{i})|\mathrm{vac}\rangle$, where $|\mathrm{vac}\rangle$ is the quantum state of the vacuum. Using the relation $\langle \mathrm{vac}|\hat{a}_{i}({\bm q'_{i}},\omega'_{i}) \hat{a}^{\dagger}_{i}({\bm q_{i}},\omega_{i})|\mathrm{vac}\rangle=\delta({\bm q'_{i}}-{\bm q_{i}})\delta(\omega_{i}-\omega'_{i})$ from quantum theory \cite{glauber1963pr}, we obtain
\begin{align}\label{vaccorfunc}
 &\langle A_{i}({\bm q'_{i}},\omega'_{i},0) A^*_{i}({\bm q_{i}},\omega_{i},0)\rangle=\frac{\hbar\omega_{i}}{4\pi\epsilon_{0}}\delta({\bm q_{i}}-{\bm q'_{i}})\delta(\omega_{i}-\omega'_{i}).
\end{align}
Thus, we find that the different modes of the classical ``vacuum'' are indeed completely uncorrelated. Moreover, the scaling factor $\hbar\omega_{i}/4\pi\epsilon_{0}$ ensures consistency with the zero-point energy of the vacuum fluctuations. The above derivation of Eq.~(\ref{vaccorfunc}) is the only calculation in this paper that explicitly appeals to quantum theory \cite{boyer}. 

We now substitute Eq.~(\ref{vaccorfunc}) in Eq.~(\ref{lgspdc1}), replace the slowly-varying term $\omega_{i}$ by its central value $\omega_{i0}$, and take it outside the integral to obtain
\begin{align}\notag
 &\langle A_{s}({\bm q_{s}},\omega_{s},L)A^*_{s}({\bm q'_{s}},\omega'_{s},L) \rangle=\frac{\hbar\omega_{i0}d_{\rm eff}^2 L^2\omega^2_{s}{\omega'_{s}}^2}{4\pi\epsilon_{0}k_{sz}k'_{sz}c^4}\\\notag&\times\iint\mathrm{d}\omega_{i}\,\mathrm{d}{\bm q_{i}}\langle A_{p}({\bm q_{s}}+{\bm q_{i}},\omega_{s}+\omega_{i}) A^*_{p}({\bm q'_{s}}+{\bm q_{i}},\omega'_{s}+\omega_{i})\rangle\\\label{lgspdc2}&\hspace{3mm}\times\mathrm{sinc}\left(\Delta k_{z}L/2\right)\mathrm{sinc}\left(\Delta k'_{z}L/2\right)\,e^{i\left(\Delta k_{z}-\Delta k'_{z}\right)L/2}.
\end{align}
The above expression is consistent with the reduced density matrix of the signal photon obtained from the full quantum-mechanical treatment of low-gain SPDC \cite{hong1985pra,grice1997pra,keller1997pra}. The integration over the classical ``vacuum'' modes on the right-hand side of Eq.~(\ref{lgspdc2}) is formally equivalent to performing a partial trace of the two-photon density matrix over the idler photon. Thus, the generated field from DFG in the low-gain regime replicates the signal field from SPDC with respect to second-order spatiotemporal correlations, and in this context the classical ``vacuum'' field in DFG plays a role equivalent to that of the idler field from SPDC. However, we emphasize that the latter equivalence is valid \textit{only} for the specific purpose of deriving the second-order correlations of the signal field because the correlations of the classical ``vacuum'' do \textit{not} reflect those of the idler field in reality.
\subsection{High-gain regime}
When the pump power is high and consequently, the nonlinear interaction is strong, we can no longer ignore the longitudinal growth of the seeding classical ``vacuum''. As a result, the derivation of the signal field correlations is slightly more involved and proceeds a bit differently. In particular, we will now use the position-time representation $V_{p}({\bm \rho},t)$ of the pump amplitude, which is related to the Fourier amplitude $A_{p}({\bm q_{p}},\omega_{p})$ as
\begin{align}\label{pumprhot}
 V_{p}({\bm \rho},t)=\iint \mathrm{d}{\bm q_{p}}\mathrm{d}\omega_{p}\,A_{p}({\bm q_{p}},\omega_{p})e^{i({\bm q_{p}}\cdot{\bm \rho}-\omega_{p}t)}. 
\end{align}
We invert the Fourier relation in Eq.~(\ref{pumprhot}), substitute it in Eq.~(\ref{dfgeqs}), and differentiate with respect to $z$ to obtain
\begin{align}\notag
 &\frac{\partial^2A_{s}({\bm q_{s}},\omega_{s},z)}{\partial z^2}=\frac{2id_{\rm eff}\omega^2_{s}}{(2\pi)^3k_{sz}c^2}\iiiint \mathrm{d}\omega_{i}\,\mathrm{d}{\bm q_{i}}\,\mathrm{d}{\bm \rho}\,\mathrm{d}t\\\notag&\hspace{1cm}\times V_{p}({\bm \rho},t)e^{-i({\bm q_{p}}\cdot{\bm \rho}-\omega_{p}t)}\,e^{i\Delta k_{z}z}\Bigg[\frac{\partial A^*_{i}({\bm q_{i}},\omega_{i},z)}{\partial z}\Bigg.\\\label{2pde1}&\hspace{4cm}\Bigg.+i\Delta k_{z}A^*_{i}({\bm q_{i}},\omega_{i},z)\Bigg].
\end{align}
In most experiments, the angular and frequency bandwidths $\delta {\bm q_{p}}$ and $\delta\omega_{p}$ of the pump are much smaller than the corresponding bandwidths $\delta {\bm q_{s}}$ and $\delta\omega_{s}$ of the generated SPDC field, respectively. Under this condition of a ``narrow-band pump'', the variation of $\Delta k_{z}$ with respect to ${\bm q_{p}}$ and $\omega_{p}$ is much slower than that of $A_{p}({\bm q_{p}},\omega_{p})$, which is usually sharply-peaked around the central wavevector ${\bm q_{p0}}=0$ and the central pump frequency $\omega_{p0}$. As a result, one can define $\Delta \bar{k}_{z}$ as the central value of $\Delta k_{z}$ evaluated for the conditions ${\bm q_{s}}+{\bm q_{i}}=0$ and $\omega_{s}+\omega_{i}=\omega_{p0}$, approximate $\Delta k_{z}\approx \Delta \bar{k}_{z}$ (this approximation is exactly true only for the monochromatic plane-wave pump case) in the second term on the right hand side, and take it outside the integral. We note that any quantity appearing with an overbar notation in the rest of this paper must be understood as the central value of that quantity evaluated for the aforementioned conditions. Using equations (\ref{dfgeqs}) and (\ref{pumprhot}) in Eq.~(\ref{2pde1}) (see Appendix B.I for details), we obtain 
\begin{align}\notag
 &\frac{\partial^2 A_{s}({\bm q_{s}},\omega_{s},z)}{\partial z^2}-i\Delta \bar{k}_{z}\frac{\partial A_{s}({\bm q_{s}},\omega_{s},z)}{\partial z}\\\label{2pde2}&\hspace{3cm}-\bar{G}^2({\bm \rho},t)\,A_{s}({\bm q_{s}},\omega_{s},z)=0,
\end{align}
where we have defined
\begin{align}\label{grhot}
 G^2({\bm \rho},t)=\frac{4d_{\rm eff}^2\omega^2_{s}\omega^2_{i}}{k_{sz}k_{iz}c^4}|V_{p}({\bm \rho},t)|^2.
\end{align}
Upon solving Eq.~(\ref{2pde2}) subject to the initial conditions $A_{s}({\bm q_{s}},\omega_{s},z=0)=0$ and $\partial A_{s}({\bm q_{s}},\omega_{s},z)/\partial z|_{z=0}$ evaluated using equations (\ref{dfg1}) and (\ref{pumprhot}), we obtain (see Appendix B.II for details)
\begin{align}\notag
&A_{s}({\bm q_{s}},\omega_{s},L)=\frac{2id_{\rm eff}\omega^2_{s}}{(2\pi)^3k_{sz}c^2}\iiiint \mathrm{d}\omega_{i}\,\mathrm{d}{\bm q_{i}}\,\mathrm{d}{\bm \rho}\,\mathrm{d}t\,V_{p}({\bm \rho},t)\\\notag&\hspace{1cm}\times e^{-i({\bm q_{p}}\cdot{\bm \rho}-\omega_{p}t)}\,A^*_{i}({\bm q_{i}},\omega_{i},0)\left[\frac{\mathrm{sinh}\,\Gamma(\Delta \bar{k}_{z},{\bm \rho},t)L}{\Gamma(\Delta \bar{k}_{z},{\bm \rho},t)}\right]\\\label{hgpdcfield}&\hspace{4cm}e^{i(\Delta k_{z}-\Delta \bar{k}_{z}/2)L},
\end{align}
where
\begin{align}\label{gamdef}
 \Gamma(\Delta \bar{k}_{z},{\bm \rho},t)\equiv\sqrt{\bar{G}^2({\bm \rho},t)-\left(\frac{\Delta \bar{k}_{z}}{2}\right)^2}.
\end{align}
Using equations (\ref{vaccorfunc}) and (\ref{hgpdcfield}), and taking $\omega_{i}\approx \omega_{i0}$ outside the integral, we find (see Appendix B.III for the detailed calculation)
\begin{align}\notag
 &\langle A_{s}({\bm q_{s}},\omega_{s},L)A^*_{s}({\bm q'_{s}},\omega'_{s},L)\rangle=\frac{2\hbar\omega_{i0}d^2_{\rm eff}\omega^2_{s}{\omega'_{s}}^2}{(2\pi)^4\epsilon_{0}k_{sz}k'_{sz}c^4}\\\notag&\hspace{5mm}\times \iint\mathrm{d}{\bm \rho}\,\mathrm{d}t\,\langle|V_{p}({\bm \rho},t)|^2\rangle\,e^{-i[({\bm q_{s}}-{\bm q'_{s}}){\bm \rho}-(\omega_{s}-\omega'_{s})t]}\\\notag&\hspace{1cm}\times\left[\frac{\mathrm{sinh}\,\Gamma(\Delta \bar{k}_{z},{\bm \rho},t)L}{\Gamma(\Delta \bar{k}_{z},{\bm \rho},t)}\right]\left[\frac{\mathrm{sinh}\,\Gamma(\Delta \bar{k'}_{z},{\bm \rho},t)L}{\Gamma(\Delta \bar{k'}_{z},{\bm \rho},t)}\right]\\\label{hgpdccorrfunc}&\hspace{4cm}\times e^{i(\Delta \bar{k}_{z}-\Delta \bar{k}'_{z})L/2}.
\end{align}
The above expression quantifies the second-order spatiotemporal correlations of the generated field for a narrow-band pump in the high-gain regime. It may be verified that Eq.~(\ref{hgpdccorrfunc}) is consistent with previous \textit{quantum} derivations of the high-gain SPDC field correlations for a monochromatic plane-wave pump \cite{klyshko1988} and a narrow-band pump \cite{brambilla2004pra} (also, see Eq.~B.1 in Appendix B of Ref.~\cite{averchenko2020pra}, where the expression derived in Ref.~\cite{brambilla2004pra} has been used for quantifying spatial correlations in degenerate SPDC with a quasimonochromatic pump with narrow angular bandwidth). We however note that owing to the ``narrow-band'' approximation involved in the above derivation, in the low-gain limit $G({\bm \rho},t)\to 0$ wherein $\Gamma(\Delta \bar{k}_{z},{\bm \rho},t)\to i\Delta \bar{k}_{z}/2$, Eq.~(\ref{hgpdccorrfunc}) is in general only \textit{approximately} equivalent to Eq.~(\ref{lgspdc2}). Nevertheless, it can be verified that the equivalence is exact for the case of a monochromatic plane-wave pump, for which the ``narrow-band'' approximation is exactly true. In the following section, we experimentally test the predictions of Eq.~(\ref{hgpdccorrfunc}) by measuring the far-field intensity profile, OAM spectrum, and frequency spectrum in high-gain SPDC for increasing gain with a pulsed Gaussian pump.
\section{Single-crystal setup: Experiments and theory}\label{single-crystal}
\begin{figure}[t!]
\centering
\includegraphics[width=85mm,keepaspectratio]{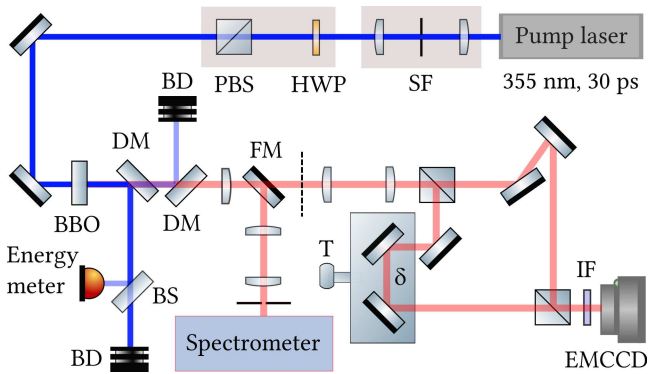}\vspace{-2mm}
\caption{Schematic of the experimental setup used for measuring the far-field intensity profile, OAM spectrum, and wavelength spectrum in high-gain SPDC for increasing pump strengths. SF: spatial filter, HWP+PBS: half-wave plate + polarizing beam-splitter combination for pump amplitude control, BBO: $\beta$-barium borate nonlinear crystal, DM: dichroic mirror, BD: beam dump, BS: beam-splitter, FM: flip mirror, T: translation stage to control the interferometric phase $\delta$, IF: interference filter centered at $710$ nm.} \vspace{-2mm}
\label{spdcsetup}
\end{figure}

 In our experimental setup depicted in Fig.~\ref{spdcsetup}, a $355$ nm vertically-polarized $30$-ps $50$-Hz pulsed Gaussian Nd:YAG laser (EKSPLA PL2231) is spatially-filtered and used to pump a $3$-mm type-I $\beta$-barium borate SPDC crystal (cut to produce horizontally-polarized collinear degenerate emission for perpendicular incidence of a vertically-polarized pump) placed at the pump's waist plane. The combination of the half-wave plate (HWP) and polarizing beam-splitter (PBS) is used to control the pump amplitude reaching the crystal. This pump amplitude is inferred up to an overall scaling factor from energy measurements using the energy meter (Coherent EnergyMax USB-J-10MB-HE). The beam-waist size $w_{p}$, defined as the $1/e^2$ half-width of the intensity profile at the waist plane, was measured using a Gentec Beamage-3.0 beam profiler to be $w_{p}=185\,\mu$m. The residual pump after the crystal is removed by means of two dichroic mirrors (DMs), and the generated SPDC field is guided towards different parts of the setup for measurements as described in the following subsections.
 
\subsection{Far-field intensity}\label{ffintsubsec}
We place a lens with its focal plane coinciding with the crystal output face, and directly image the SPDC far-field on an Andor Ixon-897 EMCCD camera (pixel size $16\,\mu$m $\times16\,\mu$m) that is placed at the plane indicated by the dashed line after the flip mirror (FM) location depicted in Fig.~\ref{spdcsetup}. The crystal orientation was approximately set to satisfy the collinear emission, and a $10$ nm interference filter (IF) centered at the degenerate wavelength $\lambda_{s}=2\lambda_{p0}=710$ nm was placed at the camera entrance. We rotate the HWP across different angles and acquire five sets of ten images each of the SPDC far-field for each corresponding pump amplitude, which is simultaneously measured using the energy meter up to an overall scaling factor. We then take diametric slices along the horizontal and vertical axes of each SPDC profile and compute the full-widths-at-half-maximum (FWHMs). We observe a slight asymmetry between the widths in the two directions due to spatial walk-off. However, we ignore walk-off effects in our study and consider the mean of the two widths as the width of the far-field profile. The FWHM width is converted from number of pixels to angle in milliradians by using the camera pixel-size and the focal length of the collimating lens. 

For obtaining the theoretical prediction, we fix the variables in the temporal degree of freedom by assuming degenerate SPDC $\lambda_{s}=710$ nm with a quasimonochromatic pump of wavelength $\lambda_{p0}=355$ nm. We verify that the pump angular bandwidth $\delta{\bm q_{p}}=2/w^2_{p}\approx 58\,\mathrm{mm}^{-2}$ is indeed much smaller than the SPDC angular bandwidth $\delta {\bm q_{s}}=|{\bm k_{s}}|/L\approx4910 \,\mathrm{mm}^{-2}$ \cite{brambilla2004pra}. We then use Eq.~(\ref{hgpdccorrfunc}) to compute the far-field intensity as
\begin{align}\notag
 \langle |A_{s}({\bm q_{s}})|^2\rangle&=\frac{K_{\rm arb}}{k^2_{sz}}\int\mathrm{d}{\bm \rho}\,\langle|V_{p}({\bm \rho})|^2\rangle\\\label{ffint}&\hspace{1cm}\times\left|\frac{\mathrm{sinh}\,\Gamma(\Delta \bar{k}_{z},{\bm \rho})L}{\Gamma(\Delta \bar{k}_{z},{\bm \rho})}\right|^2,
\end{align}
where $K_{\rm arb}$ is an overall scaling constant. Throughout this paper, the quantity $K_{\rm arb}$ must be interpreted as an arbitrary scaling factor whose value is irrelevant for our purposes. The frequency-time coordinates have been suppressed for brevity. 

For our experiments, the pump field profile and phase-mismatch are given by 
\begin{subequations}\label{ffparams}
\begin{align}
V_{p}({\bm \rho})&=g\exp\{-|{\bm \rho}|^2/w^2_{p}\},\\
\Delta \bar{k}_{z}&=|{\bm k_{p}}|-2\sqrt{|{\bm k_{s}}|^2 -|{\bm q_{s}}|^2},
\end{align}
\end{subequations}
where $g$ is a pump amplitude scaling factor, $|{\bm k_{s}}|=2\pi n_{so}/\lambda_{s}$, $|{\bm k_{p}}|=2\pi\eta_{p}(\theta_{p})/\lambda_{p0}$, $\theta_{p}$ is the angle between the pump propagation direction and the optic axis inside the crystal, and $\eta_{p}(\theta_{p})=n_{pe}n_{po}/\sqrt{n_{po}^2\sin^2\theta_{p}+n_{pe}^2\cos^2\theta_{p}}$ is the effective refractive index of the extraordinary-polarized pump inside the crystal \cite{walborn2010physrep}. The values $n_{p(e)o}$ and $n_{so}$ of the (extra)ordinary and ordinary refractive indices of BBO for the pump and signal wavelengths, respectively, can be obtained using the Sellmeier relations \cite{eimerl1987jap}
\begin{subequations}\label{bbodisrelations}
 \begin{align}
  n_{e}^2(\lambda)&=2.7405+\frac{0.0184}{\lambda^2-0.0179}-0.0155\lambda^2,\\
  n_{o}^2(\lambda)&=2.3730+\frac{0.0128}{\lambda^2-0.0156}-0.0044\lambda^2,
 \end{align}
\end{subequations}
where $\lambda$ is the corresponding wavelength in microns. Using these parameters with Eq.~(\ref{ffint}), we perform a least-squares fit between theory and experiment of the logarithmic total far-field intensity  with $g$ as the fit parameter. The value of $\theta_{p}$ for the simulations is allowed to have a small offset from the collinear emission condition $\theta^{\rm(coll)}_{p}=32.914^{\circ}$ because in experiments, $\theta_{p}$ is set by hand to approximately satisfy collinear emission. We ignore walk-off effects and exploit the rotational symmetry of the pump and the SPDC field to significantly speed up our simulations.    
\begin{figure}[t!]
\centering
\includegraphics[width=82mm,keepaspectratio]{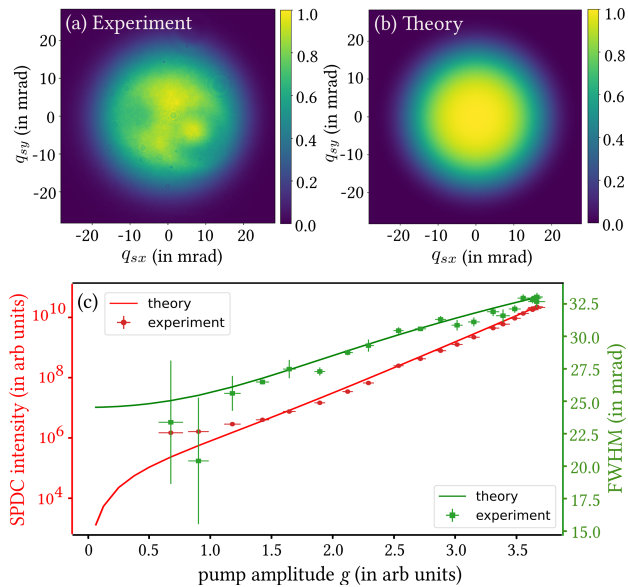}\vspace{-2mm}
\caption{Broadening of the far-field intensity profile of the SPDC field with increasing gain. (a) and (b) depict the experimental and theoretical images for $g=3.5$. The experimental and theoretical total intensities and full-widths-at-half-maximum (FWHMs) of diametric slices of such profiles for increasing pump amplitudes are depicted in (c).} \vspace{-2mm}
\label{ffintfig}
\end{figure}

We depict our experimental and theoretical results in Fig.~\ref{ffintfig}. In our simulations, we chose $\theta_{p}=32.9105^{\circ}$, which is less than $0.004^{\circ}$ away from $\theta^{\rm(coll)}_{p}=32.914^{\circ}$. The experimental and theoretical far-field intensity profiles for $g=3.5$ (in arbitrary units) are shown in Fig.~\ref{ffintfig}(a) and \ref{ffintfig}(b), respectively. Apart from some visible aberrations in the experimental image Fig.~\ref{ffintfig}(a) that appear due to burnt spots and dust on a dichroic mirror, the profiles match closely. In Fig.~\ref{ffintfig}(c), we show the trends of the full widths at half maximum (FWHMs) of diametric slices and total intensities of the far-field profiles with increasing pump amplitudes. The observed increasing trend of the FWHMs indicating the broadening of the angular spectrum with gain has also been reported previously \cite{sharapova2020prr}. Here, we find good agreement of our classical model's predictions with experiment.
\subsection{OAM spectrum}\label{oamspecsubsec}
In SPDC with a zero-OAM Gaussian pump field, OAM conservation constrains the form of the generated state to a high-dimensional superposition involving signal and idler modes with opposite anti-correlated OAM values. In the low-gain regime, these modes are populated by one photon each \cite{mair2001nature}, whereas in the high-gain regime, the same modes can be populated by higher but correlated numbers of photons \cite{beltran2017jo}. In both cases, the second-order spatial correlations of the signal field are equivalent to those of a field that is an incoherent mixture of different OAM states, and the distribution of the weights of the OAM states is referred to as the OAM spectrum. We now study the behavior of the OAM spectrum in high-gain SPDC for increasing pump strengths. 

We define the polar coordinates $(q_{s},\phi_{s})$ of ${\bm q_s}$ to be related to the Cartesian coordinates as $(q_{s}\sin \phi_{s},q_{s}\cos \phi_{s})=(q_{sx},q_{sy})$. The generated field of Eq.~(\ref{hgpdcfield}) can be written as \cite{kulkarni2017natcomm}
\begin{align}
 A_{s}(q_{s},\phi_{s})=\sum_{l=-\infty}^{+\infty}\sum_{p=0}^{\infty} \alpha_{lp} LG^{l}_{p}(q_{s})\,e^{il\phi_{s}},
\end{align}
where $l$ and $p$ are the azimuthal and radial indices of the Laguerre-Gauss mode functions $LG^{l}_{p}(q_{s})\,e^{il\phi_{s}}$, and $\alpha_{lp}$ are stochastic coefficients that satisfy $\sum_{p}\langle \alpha_{lp}\alpha^*_{l'p}\rangle=S_{l}\delta_{ll'}$, where $\delta_{ll'}$ is the Kronecker delta symbol. The quantities $S_{l}$, which include contributions from all the constituent radial modes, are collectively referred to as the OAM spectrum. The angular coherence function $W(\phi_{s1},\phi_{s2})=W(\phi_{s1}-\phi_{s2})$ defined as \cite{jha2011pra,kulkarni2017natcomm}
\begin{align}\label{angcohfunc}
W(\phi_{s1}-\phi_{s2})=\int \mathrm{d}q_{s}\,q_{s}\,\langle A_{s}(q_{s},\phi_{s1})A^*_{s}(q_{s},\phi_{s2})\rangle,
\end{align}
has been shown to be related to the OAM spectrum $S_{l}$ by the Fourier transform relation \cite{jha2011pra,kulkarni2017natcomm}
\begin{align}\label{oamspecfourier}
 S_{l}=\int_{-\pi}^{\pi} \mathrm{d}\Delta\phi\,W(\Delta\phi)\,e^{il\Delta\phi}.
\end{align}
Thus, the angular coherence function $W(\phi_{s},-\phi_{s})=W(2\phi_{s})$ can be computed using equations (\ref{hgpdccorrfunc}) and (\ref{angcohfunc}), following which Eq.~(\ref{oamspecfourier}) can be used to obtain the OAM spectrum. 
\begin{figure}[t!]
\centering
\includegraphics[width=85mm,keepaspectratio]{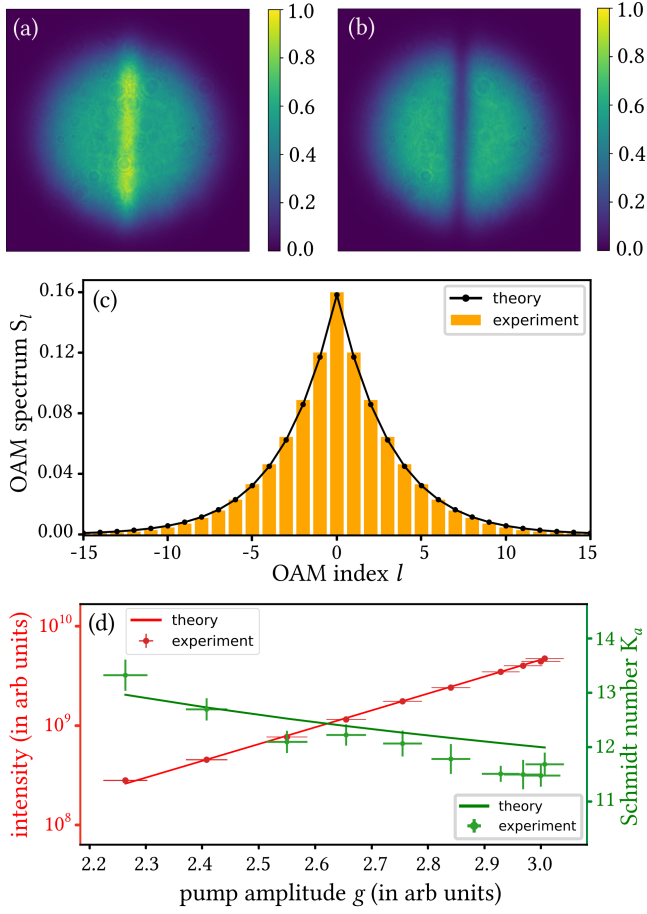}\vspace{-2mm}
\caption{Narrowing of the OAM spectrum of the SPDC signal field with increasing gain. (a) and (b) are the constructive and destructive interferograms, and (c) depicts the measured OAM spectrum alongwith the theoretical prediction for $g=3.0$. (d) depicts the total intensity and Schmidt number $K_{a}$ as a function of pump amplitude $g$.} \vspace{-2mm}
\label{oamspecfig}
\end{figure}

In our experiments, we employ the interferometric technique reported in Ref.~\cite{kulkarni2017natcomm} to measure the OAM spectrum of the high-gain SPDC field. As shown in Fig.~\ref{spdcsetup}, the SPDC field is incident into a Mach-Zehnder interferometer with an odd and even number of mirrors in the two arms, and its far-field interferograms are imaged using the EMCCD camera. The constructive and destructive interferograms are acquired by changing the interferometric phase $\delta$ using the translation stage, and their difference image is used to compute the OAM spectrum \cite{kulkarni2017natcomm}. However, this direct procedure results in large errors in the value of $S_{0}$ due to shot-to-shot fluctuations in the pump energy. We partially address this problem by averaging the interferograms over two hundred pulses, but interferometric phase drifts prevent us from increasing the acquisition time further. Moreover, the larger acquisition times also introduce errors due to systematic energy drifts of the laser. Therefore, we normalize the constructive and destructive interferograms such that the intensities in a $20\times20$ pixels region away from the central interference fringe for both images are the same. Such a normalization effectively simulates the situation where the two interferograms were obtained for identical pump energies and almost entirely eliminates the error in $S_{0}$. In addition, the sum image of the two interferograms after this normalization effectively approximates the intensity profile of the SPDC field, which is then used for fitting the experimental results to the model's predictions. In this way, we experimentally obtain the OAM spectra and the total intensities for different pump amplitudes. 

We compute the theoretical predictions using the same parameters that were used in the previous subsection. We perform a fit between theory and experiment of the logarithmic total intensities for different pump amplitudes with $g$ as the fit parameter. Using equations (\ref{hgpdccorrfunc}) and (\ref{angcohfunc}), we then compute $W(\phi_{s},-\phi_{s})$, and subsequently use Eq.~(\ref{oamspecfourier}) to compute the OAM spectrum for each pump amplitude. For our simulations, we choose $\theta_{p}=32.894^{\circ}$, which is only $0.02^{\circ}$ away from $\theta^{\rm(coll)}_{p}=32.914^{\circ}$. The rotational and reflection symmetries of the pump and SPDC fields can be used to significantly speed up the computations. 

We depict our experimental and theoretical results in Fig.~\ref{oamspecfig}. The constructive and destructive experimentally-measured interferograms for $g=3.0$ (in arb units) are depicted in Fig.~\ref{oamspecfig}(a) and Fig.~\ref{oamspecfig}(b). The measured OAM spectrum with the theoretically-predicted spectrum are shown in Fig.~\ref{oamspecfig}(c). The width of the OAM spectrum can be quantified using the angular Schmidt number $K_{a}\equiv 1/\sum_{l} S^2_{l}$, where the spectrum is normalized such that $\sum_{l}S_{l}=1$. In Fig.~\ref{oamspecfig}(d), we depict the trends of the total intensity and the Schmidt number for increasing pump amplitudes. The decreasing Schmidt number implies that the OAM spectrum narrows with increasing gain. Again, we find that our experimental measurements are in good agreement with the predictions of our classical model.
\subsection{Wavelength spectrum}\label{wavspecsubsec}
We raise the flip mirror depicted in Fig.~\ref{spdcsetup} and guide the SPDC field into a spectrometer to measure its wavelength spectrum. The spectrometer is an imaging spectrograph (Princeton Instruments Acton Series SP2558 500 mm triple-grating) with a CCD camera (PIXIS:100BR eXcelon, pixel size 20 $\mu$m $\times$ 20 $\mu$m). An aperture of 1 mm radius is used to select only the central portion of the collinear emission far-field corresponding to ${\bm q_{s}}\approx0$, and a lens with focal length $500$ mm is used to focus the light from the aperture onto the entrance slit of the spectrometer. We enable vertical binning so that the signal at a certain wavelength is the sum of the photoelectron counts over all the pixels that correspond to that wavelength. The integration time for each of the spectra is 4000 ms. We record ten spectra for each pump amplitude over the range from 610 nm to 810 nm, and to cover this range, we repeat the acquisition for different angular positions of the grating (1200 grooves per mm, 750 nm blaze). We apply a median filter to the recorded spectra to eliminate outlier peaks that appear at random locations.

For evaluating the theoretically predicted spectrum, we set ${\bm q_{s}}=0$ in collinear SPDC, and obtain the frequency spectrum from Eq.~(\ref{hgpdccorrfunc}) as
\begin{align}\notag
&\langle |A_{s}(\omega_{s})|^2\rangle=K_{\rm arb}\,\omega^2_{s}\iint\mathrm{d}{\bm \rho}\,\mathrm{d}t\,\langle|V_{p}({\bm \rho},t)|^2\rangle\,\\\label{hgfreqspec}&\hspace{3cm}\times\left|\frac{\mathrm{sinh}\,\Gamma(\Delta \bar{k}_{z},{\bm \rho},t)L}{\Gamma(\Delta \bar{k}_{z},{\bm \rho},t)}\right|^2,
\end{align}
The spatial coordinates of the signal field have been suppressed for brevity. The wavelength spectrum $S(\lambda_{s})$ can be computed using the relation $S(\lambda_{s})=\langle |A_{s}(\omega_{s})|^2\rangle |\mathrm{d}\omega_{s}/\mathrm{d}\lambda_{s}|$ that upon using $\omega_{s}=2\pi c/\lambda_{s}$ yields
\begin{align}\label{hglamspec}
 S(\lambda_{s})=K_{\rm arb}\langle |A_{s}(2\pi c/\lambda_{s})|^2\rangle/\lambda^2_{s},
\end{align}
We now compute then $S(\lambda_{s})$ using the relations,
\begin{subequations}
\begin{align}
V_{p}({\bm \rho},t)&=g\,\exp\left\{-\frac{t^2}{(2\Delta t)^2}\right\}\,\exp\left\{-\frac{|{\bm \rho}|^2}{w_{p}^2}\right\},\\
\Delta \bar{k}_{z}&=\frac{2\pi \eta_{p}(\theta_{p})}{\lambda_{p0}}-\frac{2\pi n_{o}(\lambda_{s})}{\lambda_{s}}-\frac{2\pi n_{o}(\bar{\lambda}_{i})}{\bar{\lambda}_{i}},
\end{align}
\end{subequations}
where $g$ is a pump amplitude scaling factor, $\bar{\lambda}_{i}=1/(1/\lambda_{p0}-1/\lambda_{s})$, and $\Delta t=30/2.355=12.738$ ps, where we have converted the FWHM pulse width of the power to the standard deviation width \cite{pumptransformlim}. Using these parameters and the dispersion relations (\ref{bbodisrelations}), we perform a least-squares fit of the total intensities of the recorded spectra to the total intensities predicted by equation (\ref{hgfreqspec}) with $g$ as the fit parameter. We then use the fit value of $g$ to compute the theoretically-predicted spectra for different pump amplitudes. The spatial rotational symmetry of the pump can be exploited to significantly reduce computation time. 
\begin{figure}[t!]
\centering
\includegraphics[width=85mm,keepaspectratio]{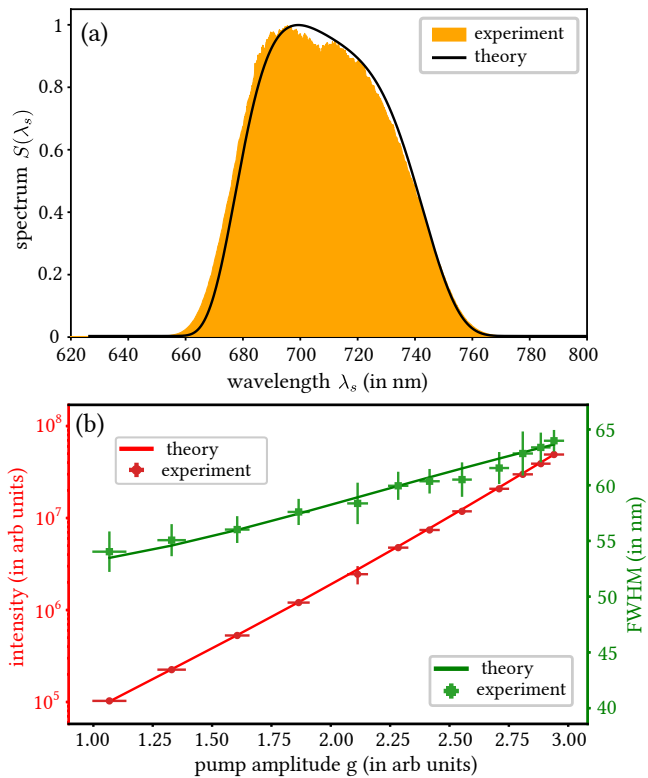}\vspace{-2mm}
\caption{Broadening of the wavelength spectrum of the SPDC signal field with increasing gain. (a) depicts the wavelength spectrum for $g=2.8$. (b) depicts the total intensity and FWHM widths of the spectra for different pump amplitudes.} \vspace{-2mm}
\label{wavspec}
\end{figure}

In Fig.~\ref{wavspec}, we depict our experimental and theoretical results. In Fig.~\ref{wavspec}(a), we depict the experimental and theoretical wavelength spectrum for $g=2.8$ (in arb units), and in Fig.~\ref{wavspec}(b), we depict the total intensities and the FWHM widths of the spectra for different pump strengths. The increasing FWHM width implies that the wavelength spectrum gets broader with increasing pump strength in high-gain SPDC \cite{spasibko2012oe}. Our experimental results are in good agreement with the theoretical predictions of our classical model. 
\section{Two-crystal setups: Theory}\label{two-crystal}
\subsection{SU(1,1) interference}
\begin{figure}[t!]
\centering
\includegraphics[width=60mm,keepaspectratio]{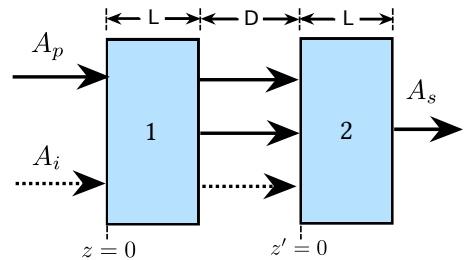}\vspace{-2mm}
\caption{Conceptual schematic depiction of SU(1,1) interference as interpreted in our classical model.} \vspace{-2mm}
\label{su11}
\end{figure}
SU(1,1) interferometers are nonlinear interferometers that can be created by essentially replacing the beam-splitters in linear SU(2) interferometers, such as Mach-Zehnder interferometers, by nonlinear crystals \cite{yurke1986pra,klyshko1993jetplett,chekhova2016aop}. It is known that such interferometers can achieve a phase sensitivity approaching the Heisenberg limit with the important advantage that they have a better loss tolerance than linear SU(2) interferometers employing squeezed states of light. In addition to phase metrology \cite{manceau2017prl}, SU(1,1) interferometers are also useful for applications such as radiation shaping \cite{lemieux2016prl}, microscopy \cite{paterova2020sciadv}, optical coherence tomography \cite{machado2020apl}, and sensing \cite{kutas2020sciadv}.

We consider the SU(1,1) setup shown in Fig.~\ref{su11} that involves two crystals 1 and 2, each of length $L$, separated by an air gap of length $D$. We define the longitudinal coordinate $z'$ such that $z'=0$ corresponds to the input face of crystal 2. We assume that there is no postselection taking place between the crystals, and that $D$ is small enough such that the far-fields of both the crystals can be assumed to overlap perfectly. As the field $A_{j}$ for $j=p,s,i$ propagates across crystal 1 and the air gap to reach crystal 2, it acquires a phase $k_{jz}L+k^{\rm (air)}_{jz}D$, where $k^{\rm (air)}_{jz}$ represents the longitudinal wavevector component of the corresponding field in air. As a result, the field generated in crystal 2 follows the relation
\begin{align}\notag
 &\frac{\partial A_{s}({\bm q_{s}},\omega_{s},z')}{\partial z'}=\frac{2id_{\rm eff}\omega^2_{s}}{k_{sz}c^2}\iint\mathrm{d}\omega_{i}\,\mathrm{d}{\bm q_{i}}\,A_{p}({\bm q_{p}},\omega_{p})\\\label{su11dfgeqn}&\hspace{1cm}\times e^{i[\Delta k^{\rm(air)}_{z}D+\Delta k_{z}L]}A^*_{i}({\bm q_{i}},\omega_{i},z')e^{i\Delta k_{z}z'},
\end{align}
where $\Delta k^{\rm(air)}_{z}=k^{\rm(air)}_{pz}-k^{\rm (air)}_{sz}-k^{\rm (air)}_{iz}$ is the longitudinal phase-mismatch in air. Using equations (\ref{dfg1}) and (\ref{su11dfgeqn}), the net field $A_{s}({\bm q_{s}},\omega_{s})$ at the output of the interferometer can be written as
\begin{align}\notag
 &A_{s}({\bm q_{s}},\omega_{s})=\frac{2id_{\rm eff}\omega^2_{s}}{k_{sz}c^2}\iint\mathrm{d}\omega_{i}\,\mathrm{d}{\bm q_{i}}\,A_{p}({\bm q_{p}},\omega_{p})\\\notag&\times\Big[\int_{0}^{L}\mathrm{d}z\,A^*_{i}({\bm q_{i}},\omega_{i},z)e^{i\Delta k_{z}z}+\,e^{i[\Delta k^{\rm(air)}_{z}D+\Delta k_{z}L]}\Big.\\&\label{su11interfeqn}\Big.\hspace{20mm}\times\int_{0}^{L}\mathrm{d}z'\,A^*_{i}({\bm q_{i}},\omega_{i},z')e^{i\Delta k_{z}z'}\Big].
\end{align}
 In what follows, we derive the second-order spatiotemporal correlations of the output field in low and high gain regimes, and illustrate the interference in the spatial domain through numerical simulations.
\subsubsection*{Low-gain regime}
 In the low-gain regime, as the interaction is weak, the longitudinal growth of the classical ``vacuum'' across both the crystals is negligible. Therefore, we approximate $A^*_{i}({\bm q_{i}},\omega_{i},z)\approx A^*_{i}({\bm q_{i}},\omega_{i},0)$ in both terms of the right-hand side of Eq.~(\ref{su11interfeqn}), and perform the integrations over $z$ and $z'$ to obtain
 \begin{align}\notag
  &A_{s}({\bm q_{s}},\omega_{s})=\frac{d_{\rm eff}L\omega^2_{s}}{k_{sz}c^2}\iint\mathrm{d}\omega_{i}\,\mathrm{d}{\bm q_{i}}\,A_{p}({\bm q_{p}},\omega_{p})A^*_{i}({\bm q_{i}},\omega_{i},0)\\\notag&\hspace{20mm}\times\mathrm{sinc}\left(\Delta k_{z}L/2\right)\,e^{i\Delta k_{z}L/2}\\&\hspace{30mm}\times\left\{1+e^{i[\Delta k^{\rm(air)}_{z}D+\Delta k_{z}L]}\right\}.
 \end{align}
 Using the above equation alongwith Eq.~(\ref{vaccorfunc}), we obtain
\begin{align}\notag
 &\langle A_{s}({\bm q_{s}},\omega_{s})A^*_{s}({\bm q'_{s}},\omega'_{s}) \rangle=\frac{\hbar\omega_{i0}d_{\rm eff}^2 L^2\omega^2_{s}{\omega'_{s}}^2}{4\pi\epsilon_{0}k_{sz}k'_{sz}c^4}\iint\mathrm{d}\omega_{i}\,\mathrm{d}{\bm q_{i}} \\\notag&\times\langle A_{p}({\bm q_{s}}+{\bm q_{i}},\omega_{s}+\omega_{i}) A^*_{p}({\bm q'_{s}}+{\bm q_{i}},\omega'_{s}+\omega_{i})\rangle\\\notag&\hspace{3mm}\times\mathrm{sinc}\left(\Delta k_{z}L/2\right)\mathrm{sinc}\left(\Delta k'_{z}L/2\right)\,e^{i\left(\Delta k_{z}-\Delta k'_{z}\right)L/2}\\\notag&\hspace{15mm}\times \left\{1+e^{i[\Delta k^{\rm(air)}_{z}D+\Delta k_{z}L]}\right\}\\\label{lgsu11corrfunc}&\hspace{30mm}\times\left\{1+e^{-i[\Delta k^{'\rm(air)}_{z}D+\Delta k'_{z}L]}\right\}.
\end{align}
The above expression quantifies the second-order spatiotemporal correlations of the output field of the SU(1,1) interferometer in the low-gain regime. 
\subsubsection*{High-gain regime}
In the high-gain regime, the initial classical ``vacuum'' $A^*_{i}({\bm q_{i}},\omega_{i},0)$ will first be amplified over the length $L$ of crystal 1 to $A^*_{i}({\bm q_{i}},\omega_{i},L)$, which then seeds the DFG process in crystal 2. In order to quantify the growth of the classical ``vacuum'', we define $A^*_{i}({\bm q_{i}},\omega_{i},z)=f(z)A^*_{i}({\bm q_{i}},\omega_{i},0)$, differentiate Eq.~(\ref{dfg2}) with respect to $z$, and make the ``narrow-band pump'' approximation to obtain
\begin{align}\label{2pdevac}
 &\frac{\partial^2 f(z)}{\partial z^2}+i\Delta \bar{k}_{z}\frac{\partial f(z)}{\partial z}-\bar{G}^2({\bm \rho},t)\,f(z)=0.
\end{align}
Solving the above Eq.~(\ref{2pdevac}) subject to the initial conditions $f(z=0)=1$ and $\partial f(z)/\partial z|_{z=0}=0$, we obtain
\begin{align}\notag
 &f(z)=e^{-i\Delta \bar{k}_{z}z/2}\Bigg\{\cosh\Gamma(\Delta \bar{k}_{z},{\bm \rho},t) z\Bigg.\\\label{vacgrowthfunc}&\hspace{2cm}\Bigg.+\frac{i\Delta \bar{k}_{z}}{2\Gamma(\Delta \bar{k}_{z},{\bm \rho},t)}\sinh\Gamma(\Delta \bar{k}_{z},{\bm \rho},t) z\Bigg\}.
\end{align}
The above relation quantifies the growth of the classical ``vacuum'' inside crystal 1. 

We now note that the generated field in crystal 2 will be governed by Eq.~(\ref{2pde2}). Moreover, the field $A_{s}({\bm q_{s}},\omega_{s},z'=0)$ at the entrance of crystal 2 is given by Eq.~(\ref{hgpdcfield}) as the field remains constant through the air gap. In addition, $\partial A_{s}({\bm q_{s}},\omega_{s},z')/\partial z' |_{z'=0}$ can be evaluated using Eq.~(\ref{su11dfgeqn}). Solving Eq.~(\ref{2pde2}) for the above initial conditions, we obtain (see Appendix B. IV for details)
\begin{align}\notag
 &A_{s}({\bm q_{s}},\omega_{s})=\frac{4id_{\rm eff}\omega^2_{s}}{(2\pi)^3k_{sz}c^2}\iiiint \mathrm{d}\omega_{i}\,\mathrm{d}{\bm q_{i}}\,\mathrm{d}{\bm \rho}\,\mathrm{d}t\,V_{p}({\bm \rho},t)\\\notag&\hspace{10mm}\times e^{-i({\bm q_{p}}\cdot{\bm \rho}-\omega_{p}t)}\,A^*_{i}({\bm q_{i}},\omega_{i},0)\left[\frac{\sinh\Gamma(\Delta \bar{k}_{z},{\bm \rho},t) L}{\Gamma(\Delta \bar{k}_{z},{\bm \rho},t)}\right]\\\label{su11hgfield}&\hspace{20mm}\times e^{i(\Delta k_{z}-\Delta \bar{k}_{z}/2)L}\,h(\Delta k^{\rm(air)}_{z},D),
\end{align}
where we have defined
\begin{align}\notag
 &h(\Delta k^{\rm(air)}_{z},D)\equiv e^{i\Delta k^{\rm(air)}_{z}D/2}\Big[\cosh\Gamma(\Delta \bar{k}_{z},{\bm \rho},t) L\Big.\\\notag&\hspace{0mm}\Big.\times \cos\left\{\Delta k^{\rm(air)}_{z}D/2\right\}-\frac{\Delta \bar{k}_{z}}{2\Gamma(\Delta \bar{k}_{z},{\bm \rho},t)}\sinh\Gamma(\Delta \bar{k}_{z},{\bm \rho},t) L\Big.\\&\hspace{35mm}\Big.\times\sin\left\{\Delta k^{\rm(air)}_{z}D/2\right\}\Big].
\end{align}
Using the above relations and Eq.~(\ref{vaccorfunc}), we obtain
\begin{align}\notag
 &\langle A_{s}({\bm q_{s}},\omega_{s})A^*_{s}({\bm q'_{s}},\omega'_{s})\rangle=\frac{4\hbar\omega_{i0}d^2_{\rm eff}\omega^2_{s}{\omega'_{s}}^2}{(2\pi)^4\epsilon_{0}k_{sz}k'_{sz}c^4}\iint\mathrm{d}{\bm \rho}\,\mathrm{d}t\\\notag&\hspace{0mm}\times \langle|V_{p}({\bm \rho},t)|^2\rangle\,e^{-i[({\bm q_{s}}-{\bm q'_{s}}){\bm \rho}-(\omega_{s}-\omega'_{s})t]}\\\notag&\hspace{0mm}\times\left[\frac{\mathrm{sinh}\,\Gamma(\Delta \bar{k}_{z},{\bm \rho},t)L}{\Gamma(\Delta \bar{k}_{z},{\bm \rho},t)}\right]\left[\frac{\mathrm{sinh}\,\Gamma(\Delta \bar{k'}_{z},{\bm \rho},t)L}{\Gamma(\Delta \bar{k'}_{z},{\bm \rho},t)}\right]\\\label{su11hgcorrfunc}&\hspace{0mm}\times \bar{h}(\Delta k^{\rm(air)}_{z},D)\,\bar{h}^*(\Delta k^{'\rm(air)}_{z},D)\,e^{i(\Delta \bar{k}_{z}-\Delta \bar{k}'_{z})L/2}.
\end{align}
The above equation quantifies the second-order spatiotemporal correlations of the output field of the SU(1,1) interferometer in the high-gain regime. 
\subsubsection*{Numerical simulations in the spatial domain}
\begin{figure}[t!]
\centering
\includegraphics[width=80mm,keepaspectratio]{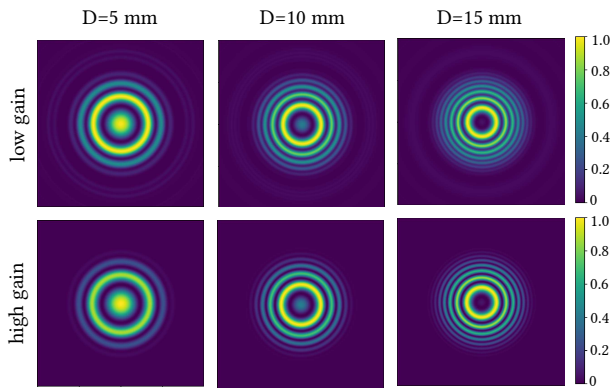}\vspace{-2mm}
\caption{SU(1,1) far-field interference patterns in the low and high gain regimes for increasing air gap lengths.} \vspace{-2mm}
\label{su11figs}
\end{figure}
We fix the frequency variables by restricting our attention to degenerate SPDC with a monochromatic pump, and suppress those variables for brevity. This restriction physically corresponds to placing a narrowband filter centered at the degenerate emission wavelength. We numerically simulate collinear degenerate SPDC for the same parameters used in Sec.~\ref{single-crystal}A. For computing the phase mismatch $\Delta k^{\rm air}_{z}$, we use the dispersion relation for air from Ref.~\cite{ciddor1996ao}, according to which the refractive index of air for the pump wavelength $355$ nm is  $n^{\rm air}_{p}=1.00028571$, and that for the signal (idler) wavelength $710$ nm is $n^{\rm air}_{s(i)}=1.00027571$. We use Eq.~(\ref{lgsu11corrfunc}) and $\Delta k_{z}=|{\bm q_{s}}-{\bm q_{i}}|^2/4k_{s}$ to compute the low-gain interference patterns, and use Eq.~(\ref{su11hgcorrfunc}) to compute the high-gain interference patterns for different air gap lengths and depict them in Fig.~\ref{su11figs}. It is evident that in both gain regimes, the interference exhibits high visibility. Moreover, with increasing air gap, the fringe separation reduces, and owing to the finite dispersion of air, the intensity at the center also varies. These predictions qualitatively agree with previous experimental observations \cite{perez2014ol, sharapova2020prr}.  
\subsection{Induced coherence}
In their seminal 1991 experiment \cite{zou1991prl}, Zou, Wang and Mandel built a modified SU(1,1) interferometric setup in which the signal fields from the two crystals were separated and superposed on a beam-splitter. Ordinarily, the signal fields from two distinct SPDC processes do not interfere with each other. However, Zou \textit{et al.} \cite{zou1991prl} observed that when the idler paths from the two processes are aligned, the signal fields become mutually coherent and exhibit interference. This intriguing effect, termed as ``induced coherence'', has not only revealed fundamental insights about interference and indistinguishability \cite{zou1991prl,wang1991pra}, but has also been harnessed in a variety of applications such as imaging \cite{lemos2014nature} and spectroscopy \cite{kalashnikov2016natphot}. 

The physical origin of induced coherence has been extensively debated over the years since its inception \cite{zou1991prl,wang1991pra,belinsky1992pla,wiseman2000pla,kolobov2017jo,lahiri2019pra}. In particular, the possibility of the coherence originating from ``induced emission'', i.e, the idler photons from the first crystal stimulating emission in the second crystal, was recognized \cite{zou1991prl, wang1991pra}. Zou \textit{et al.} ruled out this possibility by performing their experiment at weak pump power where induced emission could be shown to be negligible. They argued that the induced coherence is a consequence of the signal photon paths being rendered indistinguishable by the alignment of the idler paths. Their theoretical analysis also supported this quantum interpretation by correctly predicting the experimentally-observed linear scaling of the interference visibility with respect to the transmittance of an object placed in the idler path between the crystals. Subsequently, Wiseman and Mølmer \cite{wiseman2000pla} performed a four-mode quantum calculation for arbitrary pump strengths and showed that the linear scaling of the visibility is the true signature of induced emission being negligible, and that the scaling is \textit{not} linear when induced emission is significant. Moreover, they termed the former regime as ``quantum'', and the latter regime as ``classical''. Thus, it was implied that induced coherence without induced emission is an intrinsically quantum-mechanical effect with no classical explanation.
\begin{figure}[t!]
\centering
\includegraphics[width=75mm,keepaspectratio]{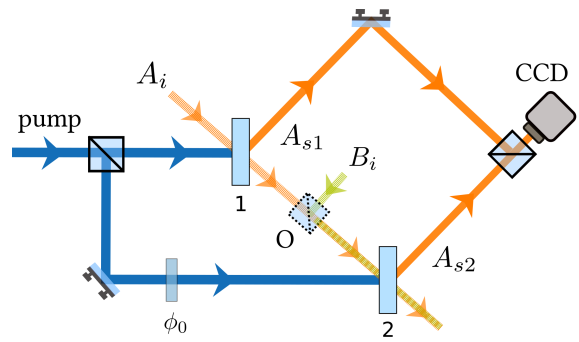}\vspace{0mm}
\caption{Conceptual schematic of the induced coherence experiment \cite{zou1991prl} interpreted in classical SPDC. Here, the partially-transmissive object O is modeled as a beam-splitter.} \vspace{0mm}
\label{inducedcoherence}
\end{figure}

In more recent years, with the series of intriguing experimental demonstrations of ``quantum imaging with undetected photons'' \cite{lemos2014nature}, ``quantifying the momentum correlation between two light beams by detecting one'' \cite{hochrainer2017pnas}, and ``interference fringes controlled by noninterfering photons'' \cite{hochrainer2017optica}, the question of whether induced coherence admits a classical explanation has gained renewed interest. While imaging with undetected light has been demonstrated in classical settings where an auxiliary external field stimulates the emission in both crystals \cite{shapiro2015scirep,cardoso2018pra}, it was explicitly stated that the case of spontaneous low-gain emission cannot be explained within classical physics \cite{cardoso2018pra}. Subsequently, a theoretical study showed that induced coherence persists even when the pump field is a single photon Fock state, where the occurrence of stimulated emission is strictly impossible \cite{lahiri2019pra}. Based on this result, the authors concluded that any classical or semiclassical explanation of induced coherence is effectively ruled out. Nevertheless, we submit that their study does not preclude the existence of a classical model that successfully captures the various induced coherence-related experiments that have been performed so far in which the pump can be treated classically. In what follows, we show that the classical SPDC model of the present paper captures several important experimentally observed features of induced coherence in low and high gain regimes, including the linear scaling of the visibility for low gain that is often regarded as the signature quantum feature of induced coherence.

In Fig.~\ref{inducedcoherence}, we depict the conceptual schematic of a prototypical induced coherence experiment involving two identical crystals as interpreted in the classical SPDC model of the present paper. The pump field is split into two equal parts: one part pumps crystal 1; the other part acquires a uniform phase $\phi_{0}$ and pumps crystal 2. A classical ``vacuum'' field $A_{i}$ seeds the process in crystal 1 and then encounters the object O that is assumed to have a complex field transmittance $Te^{i\gamma}$. The object O is modeled as a beam-splitter of splitting ratio $T:\sqrt{1-T^2}$ with a second classical ``vacuum'' $B_{i}$ that is completely uncorrelated with the first classical ``vacuum'' $A_{i}$ entering its other port. The output superposition of the two ``vacua'' from this hypothetical beam-splitter then seeds the process in crystal 2. The fields $A_{s1}$ and $A_{s2}$ generated from the two crystals 1 and 2 are superposed on a perfectly symmetric 50:50 beam-splitter and the interference is recorded on a CCD camera. We emphasize that Fig.~\ref{inducedcoherence} must be viewed as a conceptual depiction because the actual geometry of the setup can be entirely different. For instance, the experiment can also be implemented using collinear phase-matching \cite{lemos2014nature}, but the underlying concept remains the same. In what follows, we fix the temporal variables by assuming degenerate SPDC with a quasimonochromatic pump, and derive the output spatial interference patterns recorded in the induced coherence experiment in low and high gain regimes.
\subsubsection*{Low-gain regime}
\begin{figure}
\centering
\includegraphics[width=85mm,keepaspectratio]{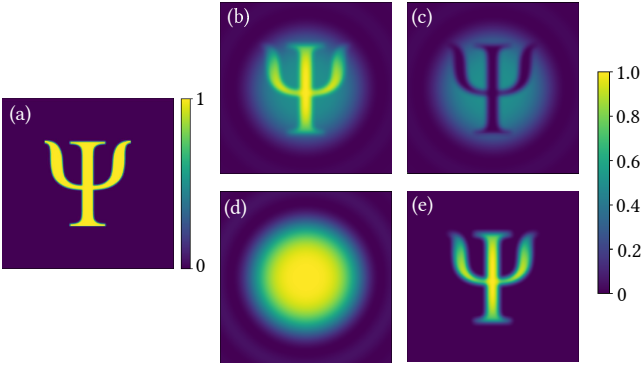}\vspace{-1mm}
\caption{Numerical simulations of the case similar to the one studied by Lemos \textit{et al.} \cite{lemos2014nature} where the object O has a binary structured transmission profile $T$. As shown in (a), we assume $T({\bm q_{i}})$ shaped in the Greek symbol $\Psi$. (b) and (c) are interferograms obtained for $\phi_{0}=0$ and $\phi_{0}=\pi$, respectively. (d) and (e) are the sum and difference images, respectively.} \vspace{-2mm}
\label{lemos}
\end{figure}
In the low-gain regime, the growth of the ``vacuum'' $A^*_{i}({\bm q_{i}},0)$ can be neglected. Using Eq.~(\ref{lgspdcfield}), the generated fields from the two crystals can then be written as
\begin{subequations}
 \begin{align}\notag
  &A_{s1}({\bm q_{s}})=\frac{K_{\rm arb}}{k_{sz}}\int\mathrm{d}{\bm q_{i}}\,A_{p}({\bm q_{p}})A^*_{i}({\bm q_{i}},0)\,\mathrm{sinc}\left(\Delta k_{z}L/2\right)\\&\hspace{40mm}\times e^{i\Delta k_{z}L/2},\\\notag
  &A_{s2}({\bm q_{s}})=\frac{K_{\rm arb}}{k_{sz}}\int\mathrm{d}{\bm q_{i}}\,A_{p}({\bm q_{p}})\,e^{i\phi_{0}}\,\mathrm{sinc}\left(\Delta k_{z}L/2\right)\\&\hspace{0mm}\times e^{i\Delta k_{z}L/2}\Big[T\,e^{i\gamma}A^*_{i}({\bm q_{i}},0)+\sqrt{1-T^2}\,B^*_{i}({\bm q_{i}},0)\Big],
 \end{align}
\end{subequations}
The above fields overlap at the beam-splitter to yield the field $A_{s}({\bm q_{s}})=\left\{A_{s1}({\bm q_{s}})+A_{s2}({\bm q_{s}})\right\}/\sqrt{2}$ at the CCD camera. We now note that $B_{i}$ has the same autocorrelations as that of $A_{i}$, i.e, $\langle B_{i}({\bm q'_{i}},0) B^*_{i}({\bm q_{i}},0)\rangle=\langle A_{i}({\bm q'_{i}},0) A^*_{i}({\bm q_{i}},0)\rangle=C\delta({\bm q_{i}}-{\bm q'_{i}})$, where $C$ is a scaling factor. However, $B_{i}$ has no mutual correlations with $A_{i}$, i.e, $\langle A_{i}({\bm q'_{i}},0) B^*_{i}({\bm q_{i}},0)\rangle=0$. Using these relations, the measured intensity takes the form
\begin{align}\notag
 I=\langle |A_{s}({\bm q_{s}})|^2\rangle&=\frac{K_{\rm arb}}{k^2_{sz}}\int\mathrm{d}{\bm q_{i}} |A_{p}({\bm q_{p}})|^2\,\mathrm{sinc}^2\left(\Delta k_{z}L/2\right)\\\label{lemoseqn}&\hspace{1cm}\times\Big\{1+T\cos\left(\gamma+\phi_{0}\right)\Big\},
\end{align}
In general, the transmittance $T$ and phase $\gamma$ could be spatially-structured, in which case they would explicitly depend on $\bm q_{i}$. However, when $T$ and $\gamma$ are uniform, the visibility $V({\bm q_{s}})=(I_{\rm max}-I_{\rm min})/(I_{\rm max}+I_{\rm min})$ scales linearly with $T$, in agreement with the quantum-mechanical prediction and experimental observations \cite{zou1991prl, wang1991pra}. Thus, our classical model is able to correctly predict the linear scaling of the visibility with object transmittance in the low-gain regime -- a feature that is often regarded as the quintessential signature of nonclassicality. We will now theoretically analyze two intriguing cases of the object O within our classical model.

We first consider the case where the object O has a binary structured transmission profile $T({\bm q_{i}})$ and constant phase $\gamma=0$ similar to the one studied by Lemos \textit{et al.} \cite{lemos2014nature}. Specifically, as shown in Fig.~\ref{lemos} (a), we assume that $T({\bm q_{i}})$ has the shape of the Greek symbol $\Psi$. We assume collinear degenerate SPDC  with the same parameters as chosen in Sec.~\ref{single-crystal}A. Using Eq.~(\ref{lemoseqn}), we numerically compute the interferograms for $\phi_{0}=0$ and $\phi_{0}=\pi$ and depict them in Fig.~\ref{lemos}(b) and Fig.~\ref{lemos}(c), respectively. The sum and difference images of these interferograms are depicted in Fig.~\ref{lemos}(d) and Fig.~\ref{lemos}(e), respectively. We note that the interferograms in Fig.~\ref{lemos}(b) and Fig.~\ref{lemos}(c) display the object O even though the interfering signal fields themselves have not directly interacted with the object, whereas the classical ``vacuum'' that has directly interacted with the object is never detected. This intriguing phenomenon was termed ``quantum imaging with undetected photons'' in Lemos \textit{et al.} \cite{lemos2014nature}, and was explained in terms of the quantum-mechanical interpretation of induced coherence. Here, it is evident that the same feature is also captured within our classical model. We emphasize that while other studies have previously demonstrated this feature in classical settings involving a real external stimulating field \cite{shapiro2015scirep,cardoso2018pra}, to our knowledge, this is the first classical treatment applied to the spontaneous low-gain limit in which Zou \textit{et al.} \cite{zou1991prl, wang1991pra} and Lemos \textit{et al.} \cite{lemos2014nature} performed their experiments.
\begin{figure}
\centering
\includegraphics[width=85mm,keepaspectratio]{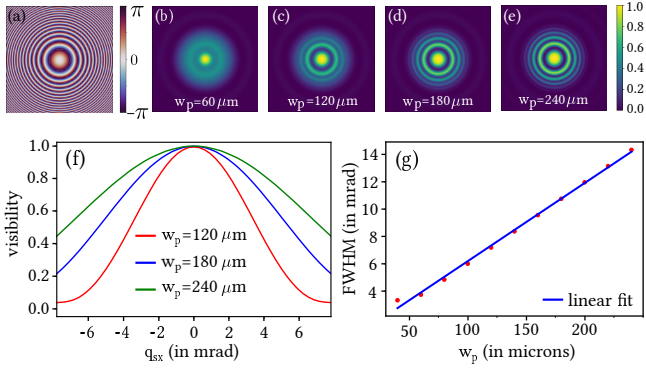}\vspace{-2mm}
\caption{Numerical low-gain simulations of the case considered in  A. Hochrainer \textit{et al.} \cite{hochrainer2017pnas} where object O is a lens with $T=1$. (a) depicts the phase $\gamma\sim |{\bm q_{i}}|^2$ modulo $2\pi$. (b), (c), (d), and (e) depict interference patterns for increasing pump beam-waist $w_{p}$. (f) depicts the visibility computed along a horizontal slice through the center of the interferograms by varying the phase $\phi_{0}$ from $0$ to $2\pi$, and (g) depicts the scaling of the FWHM of the visibility profiles with $w_{p}$.} \vspace{-2mm}
\label{hochrainer}
\end{figure}

Next, we consider the case studied by Hochrainer \textit{et al.} \cite{hochrainer2017pnas} in which the object O is a lens -- a pure phase object with $T=1$ and $\gamma=\lambda_{i}d\,|{\bm q_{i}}|^2/(4\pi)$ as depicted in Fig.~\ref{hochrainer}a. We set the equivalent free-space propagation distance $d=22.6$ mm in our analysis, and study the interference for different values of the pump beam-waist $w_{p}$. The other parameters are chosen to be the same as those used in Sec.~\ref{single-crystal}A. In Fig.~\ref{hochrainer}, we depict our numerical simulations of Eq.~(\ref{lemoseqn}) corresponding to the low-gain regime. As shown in (b), (c), (d), and (e) for small values of $w_{p}$, the interference is blurred out away from the center, whereas for large values of $w_{p}$, the interference exhibits sharp contrast even for large radial distances. This effect can be interpreted in our classical model as follows: for small $w_{p}$, the pump has a large angular bandwidth, implying a large spread in ${\bm q_{p}}$. Consequently, the far-field intensity at ${\bm q_{s}}$ is a superposition of a large number of interference patterns corresponding to several different ${\bm q_{i}}$ that satisfy ${\bm q_{i}}={\bm q_{p}}-{\bm q_{s}}$. As a result, the interference pattern is blurred. On the other hand, for large $w_{p}$, the pump's angular bandwidth is small, and consequently, the averaging effect is less pronounced leading to high interference visibility. For different values of $w_{p}$, we then compute the visibility $V({\bm q_{s}})$ along a horizontal slice through the center by numerically varying the phase $\phi_{0}$ from $0$ to $2\pi$. As shown in Fig.~\ref{hochrainer}(f), for small $w_{p}$, the visibility decays rapidly in the radial directions, whereas the decay is slower for larger $w_{p}$. In Fig.~\ref{hochrainer}(f), we plot the FWHM of the visibility curves for increasing $w_{p}$ alongwith a linear fit. This implies that the interference visibility can be used to infer the pump's angular bandwidth which, in the usual quantum mechanical picture of SPDC, also determines the transverse momentum correlations between the signal and idler photons. Therefore, the visibility can be used for ``quantifying the momentum correlations between two photons by detecting one'', as experimentally demonstrated by Hochrainer \textit{et al.} \cite{hochrainer2017pnas}. Here, we find that our above results qualitatively agree with those depicted in Fig.~2 of Ref.~\cite{hochrainer2017pnas}, but the interpretation is classical within our model.
\subsubsection*{High-gain regime}
In the high-gain regime, the initial ``vacuum'' $A^*_{i}({\bm q_{i}},0)$ is amplified in crystal 1 to $A^*_{i}({\bm q_{i}},L)=A^*_{i}({\bm q_{i}},0)f(L)$, which then superposes with the ``vacuum'' $B^*_{i}({\bm q_{i}},0)$ from the other port of O to seed crystal 2. Using equations (\ref{hgpdcfield}) and (\ref{vacgrowthfunc}), the fields from the two crystals can be written as,
\begin{subequations}\label{2sigfields}
\begin{align}\notag
 &A_{s1}({\bm q_{s}})=\frac{K_{\rm arb}}{k_{sz}}\iint\mathrm{d}{\bm q_{i}}\,\mathrm{d}{\bm \rho}\,V_{p}({\bm \rho})\,e^{-i{\bm q_{p}}\cdot{\bm \rho}}\,e^{i(\Delta k_{z}-\Delta \bar{k}_{z}/2)L}\\&\hspace{20mm}\times A^*_{i}({\bm q_{i}},0)\left[\frac{\mathrm{sinh}\,\Gamma(\Delta \bar{k}_{z},{\bm \rho})L}{\Gamma(\Delta \bar{k}_{z},{\bm \rho})}\right],\\\notag
 &A_{s2}({\bm q_{s}})=\frac{K_{\rm arb}}{k_{sz}}\iint\mathrm{d}{\bm q_{i}}\,\mathrm{d}{\bm \rho}\,V_{p}({\bm \rho})\,e^{-i({\bm q_{p}}\cdot{\bm \rho}-\phi_{0})}\\\notag&\hspace{10mm}\times e^{i(\Delta k_{z}-\Delta \bar{k}_{z}/2)L}\left[\frac{\mathrm{sinh}\,\Gamma(\Delta \bar{k}_{z},{\bm \rho})L}{\Gamma(\Delta \bar{k}_{z},{\bm \rho})}\right]\\&\hspace{5mm}\times\Big[T\,e^{i\gamma}A^*_{i}({\bm q_{i}},0)f(L)+\sqrt{1-T^2}\,B^*_{i}({\bm q_{i}},0)\Big].
\end{align}
\end{subequations}
The above fields overlap at the beam-splitter to yield the field $A_{s}({\bm q_{s}})=\left\{A_{s1}({\bm q_{s}})+A_{s2}({\bm q_{s}})\right\}/\sqrt{2}$, and the measured intensity at the CCD camera takes the form
\begin{align}\notag
&\langle |A_{s}({\bm q_{s}})|^2\rangle=\frac{K_{\rm arb}}{k^2_{sz}}\int\mathrm{d}{\bm \rho}\,\langle|V_{p}({\bm \rho})|^2\rangle\left|\frac{\mathrm{sinh}\,\Gamma(\Delta \bar{k}_{z},{\bm \rho})L}{\Gamma(\Delta \bar{k}_{z},{\bm \rho})}\right|^2\\\notag&\hspace{5mm}\times\left[1+\bar{T}^2|f(L)|^2+2\,\bar{T}\,|f(L)|\right.\\\label{mandelhgint}&\left.\hspace{10mm}\times\cos\{\mathrm{arg}\,f(L)+\bar{\gamma}+\phi_{0}\}+(1-\bar{T}^2)\right],
\end{align}
As seen in the above expression, when $|f(L)|\gg 1$, the terms quadratic in $\bar{T}$ don't cancel, and as a result, the visibility $V({\bm q_{s}})$ is \textit{not} linear in $\bar{T}$. 

\begin{figure}[t]
\centering
\includegraphics[width=85mm,keepaspectratio]{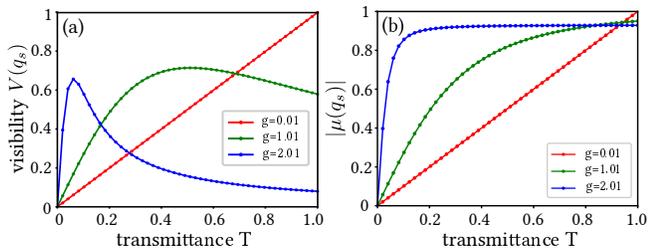}\vspace{-2mm}
\caption{(a) depicts the interference visibility, and (b) depicts the degree of coherence $|\mu({\bm q_{s}})|$ as a function of object transmittance $T$ for different pump amplitudes $g$ (in arb units).}\vspace{-2mm}
\label{finalfig}
\end{figure}

We numerically study the effect of increasing pump strength on the spatial interference. For simplicity, we now assume the object is no longer structured, but has a uniform transmittance $T$, and compute the behavior of the visibility for increasing pump strengths. We assume collinear degenerate SPDC with the same parameters chosen in Sec.~\ref{single-crystal}A. In Fig.~\ref{finalfig}(a), we depict the visibility with respect to transmittance for different pump strengths $g$ (in arb units) computed by numerically varying $\phi_{0}$ from $0$ to $2\pi$ in Eq.~(\ref{mandelhgint}). While the visibility has a linear dependence on transmittance for $g=0.01$ in the low-gain regime, the dependence is in general neither linear nor monotonic for $g=1.01$ and $g=2.01$ that correspond to the high-gain regime. The latter fact may be understood as follows: for $T=0$, the fields $A_{s1}({\bm q_{s}})$ and $A_{s2}({\bm q_{s}})$ are equal in magnitude, but have no mutual coherence, and consequently, the visibility vanishes. However, for $T>0$, the visibility increases with respect to $T$ until some critical value, but when the classical ``vacuum'' seeding the second crystal is stronger than that seeding the first crystal, the field $A_{s2}({\bm q_{s}})$ is stronger than the field $A_{s1}({\bm q_{s}})$, which causes the visibility to decrease. This behavior of the visibility in the high-gain regime has also been predicted in existing quantum calculations for a monochromatic plane-wave pump \cite{kolobov2017jo, wiseman2000pla, belinsky1992pla}. 

At this juncture, one can ask whether the decline in visibility is only due to the disproportion of the overlapping fields $A_{s1}({\bm q_{s}})$ and $A_{s2}({\bm q_{s}})$ or whether there is also a simultaneous reduction in the degree of their mutual coherence. This question is important from a practical standpoint in the context of imaging because in the former case, the fields $A_{s1}({\bm q_{s}})$ and $A_{s2}({\bm q_{s}})$ can be equalized in magnitude by suitably attenuating $A_{s2}({\bm q_{s}})$ to realize perfect visibility \cite{belinsky1992pla, kolobov2017jo}, whereas in the latter case, the visibility cannot be optimized beyond $|\mu({\bm q_{s}})|$, where the mutual degree of coherence $\mu({\bm q_{s}})$ is defined as $\mu({\bm q_{s}})\equiv \langle A^*_{s1}({\bm q_{s}})A_{s2}({\bm q_{s}})\rangle/\sqrt{\langle A_{s1}({\bm q_{s}})A^*_{s1}({\bm q_{s}})\rangle\langle A_{s2}({\bm q_{s}})A^*_{s2}({\bm q_{s}})\rangle}$. Using equations (\ref{2sigfields}), we obtain
\begin{widetext}
\begin{align}\label{degcohexp}
\mu({\bm q_{s}})=\frac{\int\mathrm{d}{\bm \rho}\,\langle|V_{p}({\bm \rho})|^2\rangle|\,\Phi(\Delta \bar{k}_{z},{\bm \rho},L)|^2\,f(L)\,\bar{T}\, e^{i\bar{\gamma}}}{\sqrt{\int\mathrm{d}{\bm \rho}\,\langle|V_{p}({\bm \rho})|^2\rangle|\,\Phi(\Delta \bar{k}_{z},{\bm \rho},L)|^2}\sqrt{\int\mathrm{d}{\bm \rho}\,\langle|V_{p}({\bm \rho})|^2\rangle|\,\Phi(\Delta \bar{k}_{z},{\bm \rho},L)|^2\left\{1+\bar{T}^2(|f(L)|^2-1)\right\}}},
\end{align}
\end{widetext}
where $\Phi(\Delta \bar{k}_{z},{\bm \rho},L)\equiv\mathrm{sinh}\,\Gamma(\Delta \bar{k}_{z},{\bm \rho},L)/\Gamma(\Delta \bar{k}_{z},{\bm \rho},L)$. It may be verified that Eq.~(\ref{degcohexp}) is consistent with previous theoretical calculations for a monochromatic plane-wave pump \cite{kolobov2017jo, wiseman2000pla}. We compute and plot the behavior of $|\mu({\bm q_{s}})|$ with respect to transmittance $T$ in Fig.~\ref{finalfig}(b) for different pump amplitudes. In much similarity with the monochromatic plane-wave pump case \cite{kolobov2017jo, wiseman2000pla}, $|\mu({\bm q_{s}})|$ scales linearly with $T$ for low-gain, but the scaling does not remain linear for increasing gain. Nevertheless, $\mu({\bm q_{s}})$ remains very high even for large gain, therefore allowing the possibility to realize imaging with high visibility by means of attenuation. There is a small discrepancy from perfect coherence in the high-gain regime for unity transmittance, i.e, $\mu({\bm q_{s}})<1$ for $T=1$. In fact, it is evident from Eq.~(\ref{degcohexp}) that owing to the expicit dependence of $f(L)$ on ${\bm \rho}$, in general $|\mu({\bm q_{s}})|\,\neq1$ for $T=1$. However, we are unable to ascertain if this discrepancy is a genuine physical effect, or an artifact of the ``narrow-band'' approximation because the discrepancy vanishes for the monochromatic plane-wave pump case.
\section{Summary and Outlook}\label{conclusions}
We describe a classical model that simulates SPDC as DFG of the pump field with a hypothetical stochastic field that mimics the effect of vacuum fluctuations. We show that the second-order spatiotemporal correlations of the field generated from DFG replicate those of the signal field from SPDC. In particular, for low gain, the second-order correlations predicted by the model are identical to those predicted by the quantum calculation of the reduced density matrix of the signal photon, whereas for high gain, we performed experimental measurements of the far-field intensity profile, OAM spectrum, and the wavelength spectrum for increasing pump strengths and demonstrated their agreement with the model's predictions. The far-field intensity profile and wavelength spectrum exhibit broadening, whereas the OAM spectrum exhibits narrowing with increasing gain. 

Next, we use the model to theoretically analyse second-order interference in SU(1,1) interferometers and induced coherence experiments. We derive analytical expressions for the second-order spatiotemporal correlations of the output field of a prototypical SU(1,1) interferometer, and illustrate some salient features of the interference in the spatial domain through numerical simulations. We then apply the model to the induced coherence experiment, and derive expressions for spatial interference in the low and high gain regimes. Interestingly, the model correctly predicts the experimentally-observed linear scaling of the visibility with object transmittance in the spontaneous low-gain limit -- a feature that is often regarded as the quintessential signature of the nonclassicality of induced coherence. We then apply the model to numerically analyze the intriguing induced coherence-related phenomena known as ``quantum imaging with undetected photons'' \cite{lemos2014nature} and ``quantifying the momentum correlation between two light beams by detecting one'' \cite{hochrainer2017pnas}. Finally, we analyze the behavior of visibility and degree of mutual coherence of the interfering fields in the high-gain regime for increasing pump strengths. We find the behavior to be consistent with previous studies that assumed a monochromatic plane-wave pump \cite{wiseman2000pla,kolobov2017jo}.  

In future, our work may potentially pave the way towards a better understanding of the classical-quantum divide in the context of SPDC and induced coherence. The model itself could be viewed as a specific application of stochastic electrodynamics -- a broader research program that attempts to explain a variety of quantum phenomena by positing the existence of a classical stochastic background radiation field that statistically mimics the zero-point vacuum fluctuations \cite{boyer1985sciam,marshall1963prsl,boyer1969pr,boyer1975prd,quantumdice2013book}. It may be possible to push the model further to investigate higher-order correlations and other intriguing fundamental effects related to SPDC and induced coherence, which could potentially shed light on the limits to which classical physics can be used to approximate the quantum world. 

In addition to fundamental implications, our work may also have significant practical applicability. Our model can be a useful theoretical tool for analysing high-gain SPDC \cite{spasibko2012oe,lemieux2016prl,beltran2017jo,sharapova2020prr} and induced coherence experiments \cite{lemos2014nature,hochrainer2017pnas,hochrainer2017optica}. Moreover, using the connection between Schmidt decomposition in quantum theory and the coherent-mode decomposition in classical coherence theory \cite{pires2011pra,kulkarni2017natcomm}, our model can be used to extract important properties of the global quantum state such as the Schmidt spectrum and Schmidt modes \cite{straupe2011pra}. A precise knowledge of these properties may not only lead to a better quantitative understanding of high-dimensional multiphoton entanglement of the high-gain SPDC field \cite{kanseri2013pra}, but may also inform experiments aimed at harnessing the underlying correlations for applications in imaging \cite{jedrkiewicz2004prl,brida2010natphot}, quantum state preparation \cite{harder2016prl},  phase metrology \cite{manceau2017prl}, radiation shaping \cite{lemieux2016prl}, microscopy \cite{paterova2020sciadv}, sensing \cite{kutas2020sciadv}, and spectroscopy \cite{kalashnikov2016natphot}.
\section*{Acknowledgments}
We acknowledge useful discussions with Samuel Lemieux, Nicol\'as Quesada, Jeff Lundeen, and Polina Sharapova. B.B acknowledges support from the Banting Postdoctoral Fellowship.
\onecolumngrid
\section*{APPENDIX A: DERIVATION OF THE DFG EQUATIONS}
\hspace{-4mm}Upon substituting equations (\ref{fourier}) and (\ref{nlpol}) in (\ref{nlowaveqn}), we obtain
\begin{align}\notag
 \left[\nabla^2+\frac{n^2_{j}\omega^2_{j}}{c^2}\right]\int\mathrm{d}{\bm q_{j}} A_{j}({\bm q_{j}},\omega_{j},z)\,e^{i({\bm q_{j}}\cdot {\bm \rho}+k_{jz}z)}&=-\frac{4d_{\rm eff}\omega^2_{j}}{c^2}\iiint \mathrm{d}\omega_{p}\,\mathrm{d}{\bm q_{p}}\,\mathrm{d}{\bm q_{l}}\, A_{p}({\bm q_{p}},\omega_{p})\,A^*_{l}({\bm q_{l}},\omega_{l},z)\\\label{appmaineqn}&\hspace{4cm}\times\,e^{i\left[({\bm q_{p}}-{\bm q_{l}})\cdot {\bm \rho}+(k_{pz}-k_{lz})z\right]},
\end{align}
for $j=s(i)$ and $l=i(s)$. We first focus only on the left hand side of the above equation. Using $\nabla\equiv(\nabla_{\perp},\partial/\partial z)$ and $|{\bm k_{j}}|=n_{j}\omega_{j}/c=\sqrt{|{\bm q_{j}}|^2+k^2_{jz}}$ and simplifying, we obtain
\begin{align*}
 &\int\mathrm{d}{\bm q_{j}}\left(|{\bm k_{j}}|^2-q^2_{j}\right)\,A_{j}({\bm q_{j}},\omega_{j},z)\,e^{i({\bm q_{j}}\cdot {\bm \rho}+k_{jz}z)}+\frac{\partial}{\partial z}\int\mathrm{d}{\bm q_{j}} \left[\frac{\partial A_{j}({\bm q_{j}},\omega_{j},z)}{\partial z}+ik_{jz}A_{j}({\bm q_{j}},\omega_{j},z)\right]\,e^{i({\bm q_{j}}\cdot {\bm \rho}+k_{jz}z)}\\
 &=\int\mathrm{d}{\bm q_{j}} \left\{k^2_{jz}\,A_{j}({\bm q_{j}},\omega_{j},z)\,e^{i({\bm q_{j}}\cdot {\bm \rho}+k_{jz}z)}+\left[\frac{\partial^2 A_{j}({\bm q_{j}},\omega_{j},z)}{\partial z^2}+2ik_{jz}\frac{\partial A_{j}({\bm q_{j}},\omega_{j},z)}{\partial z}-k^2_{jz}A_{j}({\bm q_{j}},\omega_{j},z)\right]e^{i({\bm q_{j}}\cdot {\bm \rho}+k_{jz}z)}\right\}.
\end{align*}
We make the slowly-varying envelope approximation $\frac{\partial^2}{\partial z^2}A_{j}({\bm q_{j}},\omega_{j},z)\ll k_{jz}\frac{\partial}{\partial z}A_{j}({\bm q_{j}},\omega_{j},z)$, use a different dummy variable ${\bm q'_{j}}$ instead of ${\bm q_{j}}$, and equate to the right hand side of Eq.~(\ref{appmaineqn}) to obtain
\begin{align*}
 \int\mathrm{d}{\bm q'_{j}}\,2ik'_{jz}\frac{\partial A_{j}({\bm q'_{j}},\omega_{j},z)}{\partial z}e^{i({\bm q'_{j}}\cdot {\bm \rho}+k'_{jz}z)}&=-\frac{4d_{\rm eff}\omega^2_{j}}{c^2}\iiint \mathrm{d}\omega_{p}\,\mathrm{d}{\bm q_{p}}\,\mathrm{d}{\bm q_{l}}\, A_{p}({\bm q_{p}},\omega_{p})\,A^*_{l}({\bm q_{l}},\omega_{l},z)\,e^{i\left[({\bm q_{p}}-{\bm q_{l}})\cdot {\bm \rho}+(k_{pz}-k_{lz})z\right]}.
\end{align*}
Multiplying both sides by $e^{-i{\bm q_{j}}\cdot {\bm \rho}}$ and integrating with respect to ${\bm \rho}$ over the transverse extent of the crystal, we obtain
\begin{align*}
 \iint \mathrm{d}{\bm \rho}\,\mathrm{d}{\bm q'_{j}}\,2ik'_{jz}\frac{\partial A_{j}({\bm q'_{j}},\omega_{j},z)}{\partial z}e^{i\left[({\bm q'_{j}}-{\bm q_{j}})\cdot {\bm \rho}+k'_{jz}z\right]}&=-\frac{4d_{\rm eff}\omega^2_{j}}{c^2}\int\mathrm{d}{\bm \rho} \iiint \mathrm{d}\omega_{p}\,\mathrm{d}{\bm q_{p}}\,\mathrm{d}{\bm q_{l}}\, A_{p}({\bm q_{p}},\omega_{p})\,A^*_{l}({\bm q_{l}},\omega_{l},z)\\&\hspace{6cm}\,\times e^{i\left[({\bm q_{p}}-{\bm q_{l}}-{\bm q_{j}})\cdot {\bm \rho}+(k_{pz}-k_{lz})z\right]}.
\end{align*}
We now assume that the transverse extent of the crystal is much larger than the pump spot-size and therefore, the integration over ${\bm \rho}$ can be performed over the entire infinite range of ${\bm \rho}$ to yield
\begin{align*}
 \int \,\mathrm{d}{\bm q'_{j}}\,2ik'_{jz}\frac{\partial A_{j}({\bm q'_{j}},\omega_{j},z)}{\partial z}e^{i k'_{jz}z}\delta({\bm q'_{j}}-{\bm q_{j}})&=-\frac{4d_{\rm eff}\omega^2_{j}}{c^2}\iiint \mathrm{d}\omega_{p}\,\mathrm{d}{\bm q_{p}}\,\mathrm{d}{\bm q_{l}}\, A_{p}({\bm q_{p}},\omega_{p})\,A^*_{l}({\bm q_{l}},\omega_{l},z)\,e^{i(k_{pz}-k_{lz})z}\\&\hspace{8cm}\times\delta({\bm q_{p}}-{\bm q_{l}}-{\bm q_{j}}),
\end{align*}
where the Dirac delta relation on the right-hand side expresses conservation of transverse momentum, i.e, ${\bm q_{p}}={\bm q_{s}}+{\bm q_{i}}$. Using $\omega_{p}=\omega_{s}+\omega_{i}$ and $\Delta k_{z}=k_{pz}-k_{sz}-k_{iz}$, the above equation can be simplified to yield 
\begin{align*}
 \frac{\partial A_{j}({\bm q_{j}},\omega_{j},z)}{\partial z}&=\frac{2id_{\rm eff}\omega^2_{j}}{k_{jz}c^2}\iint\mathrm{d}\omega_{l}\,\mathrm{d}{\bm q_{l}}\,A_{p}({\bm q_{p}},\omega_{p}) A^*_{l}({\bm q_{l}},\omega_{l},z)e^{i\Delta k_{z}z},
\end{align*}
which represents Eq.~(\ref{dfg1}) for $(j,l)=(s,i)$, and Eq.~(\ref{dfg2}) for $(j,l)=(i,s)$ upon complex conjugation.
\section*{APPENDIX B: DETAILED CALCULATIONS FOR THE HIGH-GAIN REGIME:}
\subsection*{I. Derivation of the second-order differential equation (\ref{2pde2}) for the signal field}
Upon substituting Eq.~(\ref{pumprhot}) in equations (\ref{dfgeqs}), we obtain
\begin{subequations}\label{dfgeqsrhot}
\begin{align}\label{dfgeqsrhot1}
 &\frac{\partial A_{s}({\bm q_{s}},\omega_{s},z)}{\partial z}=\frac{2id_{\rm eff}\omega^2_{s}}{(2\pi)^3k_{sz}c^2}\iiiint\mathrm{d}\omega_{i}\,\mathrm{d}{\bm q_{i}}\,\mathrm{d}{\bm \rho}\,\mathrm{d}t\, V_{p}({\bm \rho},t)\,e^{-i({\bm q_{p}}\cdot{\bm \rho}-\omega_{p}t)}A^*_{i}({\bm q_{i}},\omega_{i},z)e^{i\Delta k_{z}z},\\\label{dfgeqsrhot2}
 &\frac{\partial A^*_{i}({\bm q_{i}},\omega_{i},z)}{\partial z}=\frac{-2id_{\rm eff}\omega^2_{i}}{(2\pi)^3k_{iz}c^2}\iiiint\mathrm{d}\omega_{s}\,\mathrm{d}{\bm q_{s}}\,\mathrm{d}{\bm \rho}\,\mathrm{d}t \,V^*_{p}({\bm \rho},t)\,e^{i({\bm q_{p}}\cdot{\bm \rho}-\omega_{p}t)} A_{s}({\bm q_{s}},\omega_{s},z)e^{-i\Delta k_{z}z}.
\end{align}
\end{subequations}
We now consider the right-hand side of Eq.~(\ref{2pde1})
\begin{align*}
 &\frac{2id_{\rm eff}\omega^2_{s}}{(2\pi)^3k_{sz}c^2}\iiiint \mathrm{d}\omega_{i}\,\mathrm{d}{\bm q_{i}}\,\mathrm{d}{\bm \rho}\,\mathrm{d}t V_{p}({\bm \rho},t)e^{-i({\bm q_{p}}\cdot{\bm \rho}-\omega_{p}t)}\,e^{i\Delta k_{z}z}\Bigg[\frac{\partial A^*_{i}({\bm q_{i}},\omega_{i},z)}{\partial z}+i\Delta k_{z}A^*_{i}({\bm q_{i}},\omega_{i},z)\Bigg]\\
 &=\frac{4d^2_{\rm eff}\omega^2_{s}\bar{\omega}^2_{i}}{(2\pi)^6k_{sz}\bar{k}_{iz}c^4}\iiiint \mathrm{d}\omega_{i}\,\mathrm{d}{\bm q_{i}}\,\mathrm{d}{\bm \rho}\,\mathrm{d}t \iiiint \mathrm{d}\omega'_{s}\,\mathrm{d}{\bm q'_{s}}\,\mathrm{d}{\bm \rho'}\,\mathrm{d}t'  V_{p}({\bm \rho},t)V^*_{p}({\bm \rho'},t') e^{-i[({\bm q_{p}}\cdot{\bm \rho}-{\bm q'_{p}}\cdot{\bm \rho'})-(\omega_{p}t-\omega'_{p}t')]} A_{s}({\bm q'_{s}},\omega'_{s},z)\\&\hspace{2cm}\times e^{i(\Delta k_{z}-\Delta k'_{z})z}+i\Delta \bar{k}_{z}\frac{2id_{\rm eff}\omega^2_{s}}{(2\pi)^3k_{sz}c^2}\iiiint\mathrm{d}\omega_{i}\,\mathrm{d}{\bm q_{i}}\,\mathrm{d}{\bm \rho}\,\mathrm{d}t\, V_{p}({\bm \rho},t)\,e^{-i({\bm q_{p}}\cdot{\bm \rho}-\omega_{p}t)}A^*_{i}({\bm q_{i}},\omega_{i},z)e^{i\Delta k_{z}z}\\
 &=\frac{4d^2_{\rm eff}\omega^2_{s}\bar{\omega}^2_{i}}{(2\pi)^6k_{sz}\bar{k}_{iz}c^4}\iiiint \mathrm{d}\omega_{i}\,\mathrm{d}{\bm q_{i}}\,\mathrm{d}{\bm \rho}\,\mathrm{d}t \iiiint \mathrm{d}\omega'_{s}\,\mathrm{d}{\bm q'_{s}}\,\mathrm{d}{\bm \rho'}\,\mathrm{d}t'  V_{p}({\bm \rho},t)V^*_{p}({\bm \rho'},t') e^{-i[({\bm q_{s}}\cdot{\bm \rho}-{\bm q'_{s}}\cdot{\bm \rho'})-(\omega_{s}t-\omega'_{s}t')+{\bm q_{i}}({\bm \rho}-{\bm \rho'})-\omega_{i}(t-t')]}\\&\hspace{10cm}\times A_{s}({\bm q'_{s}},\omega'_{s},z) e^{i(\Delta k_{z}-\Delta k'_{z})z}+i\Delta \bar{k}_{z}\frac{\partial A_{s}({\bm q_{s}},\omega_{s},z)}{\partial z}\\
 &=\frac{4d^2_{\rm eff}\omega^2_{s}\bar{\omega}^2_{i}}{(2\pi)^3k_{sz}\bar{k}_{iz}c^4}\iint\mathrm{d}{\bm \rho}\,\mathrm{d}t \iiiint \mathrm{d}\omega'_{s}\,\mathrm{d}{\bm q'_{s}}\,\mathrm{d}{\bm \rho'}\,\mathrm{d}t'  V_{p}({\bm \rho},t)V^*_{p}({\bm \rho'},t') e^{-i[({\bm q_{s}}\cdot{\bm \rho}-{\bm q'_{s}}\cdot{\bm \rho'})-(\omega_{s}t-\omega'_{s}t')]}\delta({\bm \rho}-{\bm \rho'})\delta(t-t') A_{s}({\bm q'_{s}},\omega'_{s},z)\\&\hspace{12cm}\times e^{i(\Delta k_{z}-\Delta k'_{z})z}+i\Delta \bar{k}_{z}\frac{\partial A_{s}({\bm q_{s}},\omega_{s},z)}{\partial z}\\
 &=\frac{4d^2_{\rm eff}\omega^2_{s}\bar{\omega}^2_{i}}{(2\pi)^3k_{sz}\bar{k}_{iz}c^4}\iint\mathrm{d}{\bm \rho}\,\mathrm{d}t \iint \mathrm{d}\omega'_{s}\,\mathrm{d}{\bm q'_{s}}\,|V_{p}({\bm \rho},t)|^2 e^{-i[({\bm q_{s}}-{\bm q'_{s}})\cdot{\bm \rho}-(\omega_{s}-\omega_{s})t]}\,A_{s}({\bm q'_{s}},\omega'_{s},z)\,e^{i(\Delta k_{z}-\Delta k'_{z})z}+i\Delta \bar{k}_{z}\frac{\partial A_{s}({\bm q_{s}},\omega_{s},z)}{\partial z}\\
 &=\frac{4d^2_{\rm eff}\omega^2_{s}\bar{\omega}^2_{i}}{k_{sz}\bar{k}_{iz}c^4}|V_{p}({\bm \rho},t)|^2\iint \mathrm{d}\omega'_{s}\,\mathrm{d}{\bm q'_{s}}\,\delta({\bm q_{s}}-{\bm q'_{s}})\,\delta(\omega_{s}-\omega_{s})\,A_{s}({\bm q'_{s}},\omega'_{s},z)\,e^{i(\Delta k_{z}-\Delta k'_{z})z}+i\Delta \bar{k}_{z}\frac{\partial A_{s}({\bm q_{s}},\omega_{s},z)}{\partial z}\\
 &=\frac{4d^2_{\rm eff}\omega^2_{s}\bar{\omega}^2_{i}}{k_{sz}\bar{k}_{iz}c^4}|V_{p}({\bm \rho},t)|^2\,A_{s}({\bm q_{s}},\omega_{s},z)+i\Delta \bar{k}_{z}\frac{\partial A_{s}({\bm q_{s}},\omega_{s},z)}{\partial z}.
\end{align*}
We note that in the second-last step of the above calculation, we have used the fact that the pump intensity profile $|V_{p}({\bm \rho},t)|^2$ has a very slow variation with respect to ${\bm \rho}$ and $t$ due to our assumption that the frequency and angular bandwidth of the pump is much smaller than that of the generated field. Consequently, the function $|V_{p}({\bm \rho},t)|^2$ could be taken out of the integral. Upon shifting the above result to the left-hand side of Eq.~(\ref{2pde1}), we obtain Eq.~(\ref{2pde2}).
\subsection*{II. Solution of the second-order differential equation (\ref{2pde2}) for the signal field}
Substituting an ansatz solution of the form $A_{s}({\bm q_{s}},\omega_{s},z)=e^{rz}$ in Eq.~(\ref{2pde2}), we obtain $r^2-i\Delta \bar{k}_{z}r-\bar{G}^2({\bm \rho},t)=0$, which has the roots $r_{\pm}=i\Delta \bar{k}_{z}/2\pm \Gamma(\Delta \bar{k}_{z},{\bm \rho},t)$, where $\Gamma(\Delta \bar{k}_{z},{\bm \rho},t)$ is defined by Eq.~(\ref{gamdef}). The general solution then takes the form $A_{s}({\bm q_{s}},\omega_{s},z)=Ce^{r_{+}z}+De^{r_{-}z}$, where $C$ and $D$ are scaling factors to be determined by initial conditions. The first condition $A_{s}({\bm q_{s}},\omega_{s},z=0)=0$ implies $C+D=0$, and the value of $\partial A_{s}({\bm q_{s}},\omega_{s},z)/\partial z|_{z=0}$ obtained by evaluating Eq.~(\ref{dfgeqsrhot1}) for $z=0$ yields
\begin{align*}
 C=\frac{id_{\rm eff}\omega^2_{s}}{(2\pi)^3 k_{sz}c^2}\iiiint\mathrm{d}\omega_{i}\,\mathrm{d}{\bm q_{i}}\,\mathrm{d}{\bm \rho}\,\mathrm{d}t\, \frac{V_{p}({\bm \rho},t)}{\Gamma(\Delta \bar{k}_{z},{\bm \rho},t)}\,e^{-i({\bm q_{p}}\cdot{\bm \rho}-\omega_{p}t)}A^*_{i}({\bm q_{i}},\omega_{i},0).
\end{align*}
We again use the approximation that the function $\Gamma(\Delta \bar{k}_{z},{\bm \rho},t)$ from its dependence on $|V_{p}({\bm \rho},t)|^2$ has a very slow variation with respect to ${\bm \rho}$ and $t$. The above relation then implies $A_{s}({\bm q_{s}},\omega_{s},z)=2C\,e^{i\Delta \bar{k}_{z}z/2}\,\mathrm{sinh}\left[\Gamma(\Delta \bar{k}_{z},{\bm \rho},t) z\right]$, which yields Eq.~(\ref{hgpdcfield}) for $z=L$. 
\subsection*{III. Evaluating the high-gain spatiotemporal correlation function of the signal field:}
Using Eq.~(\ref{hgpdcfield}), we evaluate
\begin{align*}
 &\langle A_{s}({\bm q_{s}},\omega_{s},z)A^*_{s}({\bm q'_{s}},\omega'_{s},z)\rangle=\frac{4d^2_{\rm eff}\omega^2_{s}{\omega'_{s}}^2}{(2\pi)^6k_{sz}k'_{sz}c^4}\iiiint\mathrm{d}\omega_{i}\,\mathrm{d}{\bm q_{i}}\,\mathrm{d}{\bm \rho}\,\mathrm{d}t\iiiint\mathrm{d}\omega'_{i}\,\mathrm{d}{\bm q'_{i}}\,\mathrm{d}{\bm \rho'}\,\mathrm{d}t' \langle V_{p}({\bm \rho},t)V_{p}({\bm \rho'},t')\rangle\\&\times e^{-i[({\bm q_{p}}\cdot{\bm \rho}-{\bm q'_{p}}\cdot{\bm \rho'})-(\omega_{p}t-\omega'_{p}t')]}\,\langle A^*_{i}({\bm q_{i}},\omega_{i},0)A_{i}({\bm q'_{i}},\omega'_{i},0)\rangle\left[\frac{\mathrm{sinh}\,\Gamma(\Delta \bar{k}_{z},{\bm \rho},t)L}{\Gamma(\Delta \bar{k}_{z},{\bm \rho},t)}\right]\left[\frac{\mathrm{sinh}\,\Gamma(\Delta \bar{k'}_{z},{\bm \rho'},t')L}{\Gamma(\Delta \bar{k'}_{z},{\bm \rho'},t')}\right]e^{i[(\Delta k_{z}-\Delta k'_{z})+(\Delta \bar{k}_{z}-\Delta \bar{k}'_{z})/2]L}\\
 &=\frac{4d^2_{\rm eff}\omega^2_{s}{\omega'_{s}}^2}{(2\pi)^6k_{sz}k'_{sz}c^4}\iiiint\mathrm{d}\omega_{i}\,\mathrm{d}{\bm q_{i}}\,\mathrm{d}{\bm \rho}\,\mathrm{d}t\iiiint\mathrm{d}\omega'_{i}\,\mathrm{d}{\bm q'_{i}}\,\mathrm{d}{\bm \rho'}\,\mathrm{d}t' \langle V_{p}({\bm \rho},t)V_{p}({\bm \rho'},t')\rangle\, e^{-i[({\bm q_{p}}\cdot{\bm \rho}-{\bm q'_{p}}\cdot{\bm \rho'})-(\omega_{p}t-\omega'_{p}t')]}\\&\hspace{2cm}\times\frac{\hbar\omega_{i}}{4\pi\epsilon_{0}}\delta({\bm q_{i}}-{\bm q'_{i}})\delta(\omega_{i}-\omega'_{i})\left[\frac{\mathrm{sinh}\,\Gamma(\Delta \bar{k}_{z},{\bm \rho},t)L}{\Gamma(\Delta \bar{k}_{z},{\bm \rho},t)}\right]\left[\frac{\mathrm{sinh}\,\Gamma(\Delta \bar{k'}_{z},{\bm \rho'},t')L}{\Gamma(\Delta \bar{k'}_{z},{\bm \rho'},t')}\right]e^{i[(\Delta k_{z}-\Delta k'_{z})+(\Delta \bar{k}_{z}-\Delta \bar{k}'_{z})/2]L}\\
 &=\frac{4d^2_{\rm eff}\omega^2_{s}{\omega'_{s}}^2}{(2\pi)^6k_{sz}k'_{sz}c^4}\iiiint\mathrm{d}\omega_{i}\,\mathrm{d}{\bm q_{i}}\,\mathrm{d}{\bm \rho}\,\mathrm{d}t \frac{\hbar\omega_{i}}{4\pi\epsilon_{0}}\iint\mathrm{d}{\bm \rho'}\,\mathrm{d}t' \langle V_{p}({\bm \rho},t)V_{p}({\bm \rho'},t')\rangle\, e^{-i[({\bm q_{s}}\cdot{\bm \rho}-{\bm q'_{s}}\cdot{\bm \rho'})-(\omega_{s}t-\omega'_{s}t')]}e^{i[{\bm q_{i}}\cdot({\bm \rho}-{\bm \rho'})-\omega_{i}(t-t')]}\\&\hspace{6cm}\times \left[\frac{\mathrm{sinh}\,\Gamma(\Delta \bar{k}_{z},{\bm \rho},t)L}{\Gamma(\Delta \bar{k}_{z},{\bm \rho},t)}\right]\left[\frac{\mathrm{sinh}\,\Gamma(\Delta \bar{k'}_{z},{\bm \rho'},t')L}{\Gamma(\Delta \bar{k'}_{z},{\bm \rho'},t')}\right]e^{i[(\Delta k_{z}-\Delta k'_{z})+(\Delta \bar{k}_{z}-\Delta \bar{k}'_{z})/2]L}\\
 &=\frac{2\hbar\omega_{i0}d^2_{\rm eff}\omega^2_{s}{\omega'_{s}}^2}{(2\pi)^4\epsilon_{0}k_{sz}k'_{sz}c^4}\iint\mathrm{d}{\bm \rho}\,\mathrm{d}t\iint\mathrm{d}{\bm \rho'}\,\mathrm{d}t' \langle V_{p}({\bm \rho},t)V_{p}({\bm \rho'},t')\rangle\, e^{-i[({\bm q_{s}}\cdot{\bm \rho}-{\bm q'_{s}}\cdot{\bm \rho'})-(\omega_{s}t-\omega'_{s}t')]}\delta({\bm \rho}-{\bm \rho'})\delta(t-t')\\&\hspace{8cm}\times \left[\frac{\mathrm{sinh}\,\Gamma(\Delta \bar{k}_{z},{\bm \rho},t)L}{\Gamma(\Delta \bar{k}_{z},{\bm \rho},t)}\right]\left[\frac{\mathrm{sinh}\,\Gamma(\Delta \bar{k'}_{z},{\bm \rho'},t')L}{\Gamma(\Delta \bar{k'}_{z},{\bm \rho'},t')}\right]e^{i(\Delta \bar{k}_{z}-\Delta \bar{k}'_{z})L/2}\\
 &=\frac{2\hbar\omega_{i0}d^2_{\rm eff}\omega^2_{s}{\omega'_{s}}^2}{(2\pi)^4\epsilon_{0}k_{sz}k'_{sz}c^4}\iint\mathrm{d}{\bm \rho}\,\mathrm{d}t\langle |V_{p}({\bm \rho},t)|^2\rangle\, e^{-i[({\bm q_{s}}-{\bm q'_{s}})\cdot{\bm \rho}-(\omega_{s}-\omega'_{s})t]}\left[\frac{\mathrm{sinh}\,\Gamma(\Delta \bar{k}_{z},{\bm \rho},t)L}{\Gamma(\Delta \bar{k}_{z},{\bm \rho},t)}\right]\left[\frac{\mathrm{sinh}\,\Gamma(\Delta \bar{k'}_{z},{\bm \rho},t)L}{\Gamma(\Delta \bar{k'}_{z},{\bm \rho},t)}\right]e^{i(\Delta \bar{k}_{z}-\Delta \bar{k}'_{z})L/2},
\end{align*}
which is Eq.~(\ref{hgpdccorrfunc}).
\subsection*{IV. Evaluating the output field from the SU(1,1) setup:}
As the field $A_{s}({\bm q_{s}},\omega_{s},z')$ satisfies Eq.~(\ref{2pde2}), we can assume an ansatz solution of the form $A_{s}({\bm q_{s}},\omega_{s},z')=Ce^{r_{+}z'}+De^{r_{-}z'}$, where $r_{\pm}=i\Delta \bar{k}_{z}/2\pm \Gamma(\Delta \bar{k}_{z},{\bm \rho},t)$. Thus, we have $C+D=A_{s}({\bm q_{s}},\omega_{s},z'=0)$ and $Cr_{+}+Dr_{-}=\dot{A}_{s}({\bm q_{s}},\omega_{s},z'=0)=\partial A_{s}({\bm q_{s}},\omega_{s},z')/\partial z|_{z'=0}$, which implies $C=\left\{\dot{A}_{s}({\bm q_{s}},\omega_{s},z'=0)-r_{-}A_{s}({\bm q_{s}},\omega_{s},z'=0)\right\}/(r_{+}-r_{-})$ and $D=\left\{r_{+}A_{s}({\bm q_{s}},\omega_{s},z'=0)-\dot{A}_{s}({\bm q_{s}},\omega_{s},z'=0)\right\}/(r_{+}-r_{-})$. Substituting from equations (\ref{hgpdcfield}) and (\ref{su11dfgeqn}), we obtain
\begin{align*}
 &A_{s}({\bm q_{s}},\omega_{s},z'=L)=\frac{2id_{\rm eff}\omega^2_{s}}{(2\pi)^3k_{sz}c^2}\iiiint \mathrm{d}\omega_{i}\,\mathrm{d}{\bm q_{i}}\,\mathrm{d}{\bm \rho}\,\mathrm{d}t\,V_{p}({\bm \rho},t)\,e^{-i({\bm q_{p}}\cdot{\bm \rho}-\omega_{p}t)}\,A^*_{i}({\bm q_{i}},\omega_{i},0)\,e^{i(\Delta k_{z}-\Delta \bar{k}_{z}/2)L}\\&\hspace{2cm}\times\left\{e^{\Gamma L}\left[\frac{e^{i\Delta k^{\rm(air)}_{z}D}}{2\Gamma}\Bigg\{\cosh\Gamma L+\frac{i\Delta \bar{k}_{z}}{2\Gamma}\sinh\Gamma L\Bigg\}+\left\{\Gamma-i\Delta \bar{k}_{z}/2\right\}\frac{\mathrm{sinh}\,\Gamma L}{2\Gamma^2}\right]\right.\\&\hspace{4cm}\left.+e^{-\Gamma L}\left[\left\{i\Delta \bar{k}_{z}/2+\Gamma\right\}\frac{\mathrm{sinh}\,\Gamma L}{2\Gamma^2}-\frac{e^{i\Delta k^{\rm(air)}_{z}D}}{2\Gamma}\Bigg\{\cosh\Gamma L+\frac{i\Delta \bar{k}_{z}}{2\Gamma}\sinh\Gamma L\Bigg\}\right]\right\}\\
 &=\frac{2id_{\rm eff}\omega^2_{s}}{(2\pi)^3k_{sz}c^2}\iiiint \mathrm{d}\omega_{i}\,\mathrm{d}{\bm q_{i}}\,\mathrm{d}{\bm \rho}\,\mathrm{d}t\,V_{p}({\bm \rho},t)\,e^{-i({\bm q_{p}}\cdot{\bm \rho}-\omega_{p}t)}\,A^*_{i}({\bm q_{i}},\omega_{i},0)\,e^{i(\Delta k_{z}-\Delta \bar{k}_{z}/2)L}\\&\hspace{2cm}\times\left\{e^{i\Delta k^{\rm(air)}_{z}D}\Bigg\{\frac{\sinh2\Gamma L}{2\Gamma}+\frac{i\Delta \bar{k}_{z}}{2\Gamma^2}\sinh^2\Gamma L\Bigg\}+\frac{\mathrm{sinh}\,\Gamma L}{2\Gamma^2}\left(2\Gamma\cosh \Gamma L-i\Delta \bar{k}_{z}\sinh \Gamma L\right)\right\}\\
 &=\frac{2id_{\rm eff}\omega^2_{s}}{(2\pi)^3k_{sz}c^2}\iiiint \mathrm{d}\omega_{i}\,\mathrm{d}{\bm q_{i}}\,\mathrm{d}{\bm \rho}\,\mathrm{d}t\,V_{p}({\bm \rho},t)\,e^{-i({\bm q_{p}}\cdot{\bm \rho}-\omega_{p}t)}\,A^*_{i}({\bm q_{i}},\omega_{i},0)\,e^{i(\Delta k_{z}-\Delta \bar{k}_{z}/2)L}\\&\hspace{4cm}\times\left\{\frac{\sinh2\Gamma L}{2\Gamma}\left(1+e^{i\Delta k^{\rm(air)}_{z}D}\right)-\frac{i\Delta \bar{k}_{z}}{2}\left(\frac{\sinh\Gamma L}{\Gamma}\right)^2\left(1-e^{i\Delta k^{\rm(air)}_{z}D}\right)\right\}
 \end{align*}
 \begin{align*}
 &=\frac{2id_{\rm eff}\omega^2_{s}}{(2\pi)^3k_{sz}c^2}\iiiint \mathrm{d}\omega_{i}\,\mathrm{d}{\bm q_{i}}\,\mathrm{d}{\bm \rho}\,\mathrm{d}t\,V_{p}({\bm \rho},t)\,e^{-i({\bm q_{p}}\cdot{\bm \rho}-\omega_{p}t)}\,A^*_{i}({\bm q_{i}},\omega_{i},0)\,e^{i(\Delta k_{z}-\Delta \bar{k}_{z}/2)L}\,e^{i\Delta k^{\rm(air)}_{z}D/2}\\&\hspace{4cm}\times\left\{\frac{\sinh2\Gamma L}{\Gamma}\cos\left\{\Delta k^{\rm(air)}_{z}D/2\right\}-\Delta \bar{k}_{z}\left(\frac{\sinh\Gamma L}{\Gamma}\right)^2\sin\left\{\Delta k^{\rm(air)}_{z}D/2\right\}\right\}\\
 &=\frac{4id_{\rm eff}\omega^2_{s}}{(2\pi)^3k_{sz}c^2}\iiiint \mathrm{d}\omega_{i}\,\mathrm{d}{\bm q_{i}}\,\mathrm{d}{\bm \rho}\,\mathrm{d}t\,V_{p}({\bm \rho},t)\,e^{-i({\bm q_{p}}\cdot{\bm \rho}-\omega_{p}t)}\,A^*_{i}({\bm q_{i}},\omega_{i},0)\,e^{i(\Delta k_{z}-\Delta \bar{k}_{z}/2)L}\,e^{i\Delta k^{\rm(air)}_{z}D/2}\left(\frac{\sinh\Gamma L}{\Gamma}\right)\\&\hspace{4cm}\times\left[\cosh\Gamma L\,\cos\left\{\Delta k^{\rm(air)}_{z}D/2\right\}-\frac{\Delta \bar{k}_{z}}{2\Gamma}\sinh\Gamma L\sin\left\{\Delta k^{\rm(air)}_{z}D/2\right\}\right],
\end{align*}
which is Eq.~(\ref{su11hgfield}).
\twocolumngrid
\bibliography{classicalspdc}

\begin{thebibliography}{78}
\bibitem [{pum()}]{pumptransformlim}%
  \BibitemOpen
  \href@noop {} {}\bibinfo {note} {We have implicitly assumed that the pump
  field is a transform-limited Gaussian pulse, which is not strictly true.
  However, for our purposes, this assumption will not introduce significant
  errors.}\BibitemShut {Stop}%
\end{thebibliography}
\end{document}